\documentclass[twocolumn]{aastex62}
\usepackage{bm}
\begin{document}

\title{Seasonal structures in Saturn's dusty Roche Division correspond to periodicities of the planet's magnetosphere}

\author[0000-0002-7867-7674]{R.O. Chancia}
\affiliation{Department of Physics, University of Idaho, Moscow, Idaho, USA}

\author[0000-0002-8592-0812]{M.M. Hedman}
\affiliation{Department of Physics, University of Idaho, Moscow, Idaho, USA}

\author[0000-0002-4041-0034]{S.W.H. Cowley}
\affiliation{Department of Physics and Astronomy, University of Leicester, Leicester, UK}

\author[0000-0001-7442-4154]{G. Provan}
\affiliation{Department of Physics and Astronomy, University of Leicester, Leicester, UK}

\author[0000-0002-3064-1082]{S.-Y. Ye}
\affiliation{Department of Physics and Astronomy, University of Iowa, Iowa City, Iowa, USA}

\correspondingauthor{Robert O. Chancia}
\email{rchancia@uidaho.edu}

\begin{abstract}
We identify multiple periodic dusty structures in Saturn's Roche Division, a faint region spanning the $\sim3000$ km between the A and F rings. The locations and extent of these features vary over \textit{Cassini's} tour of the Saturn system, being visible in 2006 and 2016-2017, but not in 2012-2014. These changes can be correlated with variations in Saturn's magnetospheric periods. 
In 2006 and 2016-2017, one of the drifting magnetospheric periods would produce a 3:4 resonance within the Roche Division, but in 2012-2014 these resonances would move into the A ring as the magnetospheric periods converged. A simple model of magnetic perturbations indicates that the magnetic field oscillations responsible for these structures have amplitudes of a few nanotesla, comparable to the magnetic field oscillation amplitudes of planetary period oscillations measured by the magnetometer onboard \textit{Cassini}. However, some previously unnoticed features at higher radii have expected pattern speeds that are much slower than the magnetospheric periodicities. These structures may reflect an unexpectedly long-range propagation of resonant perturbations within dusty rings. 
\end{abstract}

\keywords{Planetary rings; Saturn, rings; Saturn, magnetosphere; Resonances, rings}

\section{Introduction \label{intro}}
\begin{figure*}[t]
\centering
\includegraphics[width=.95\linewidth]{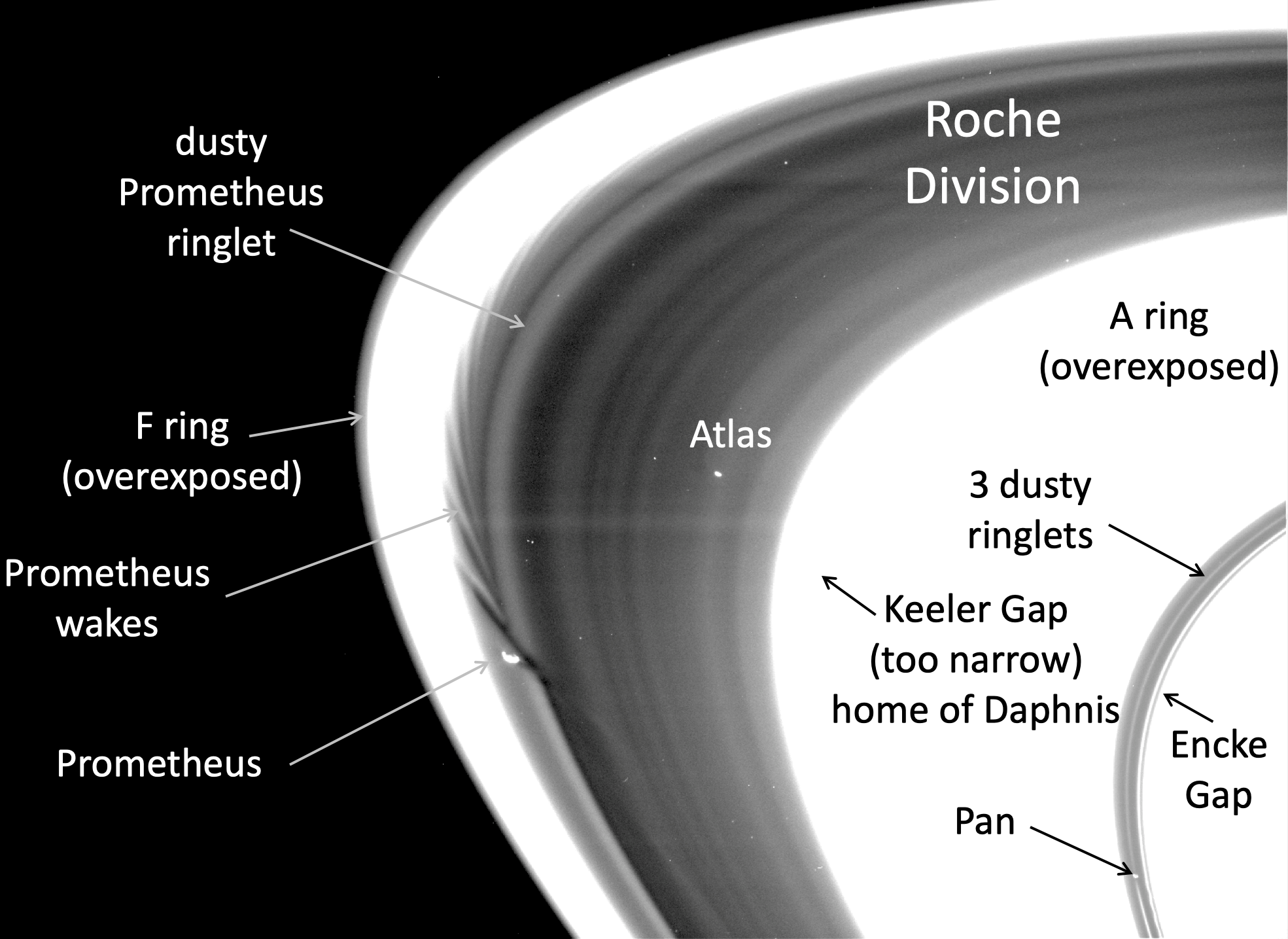}
\includegraphics[width=\linewidth]{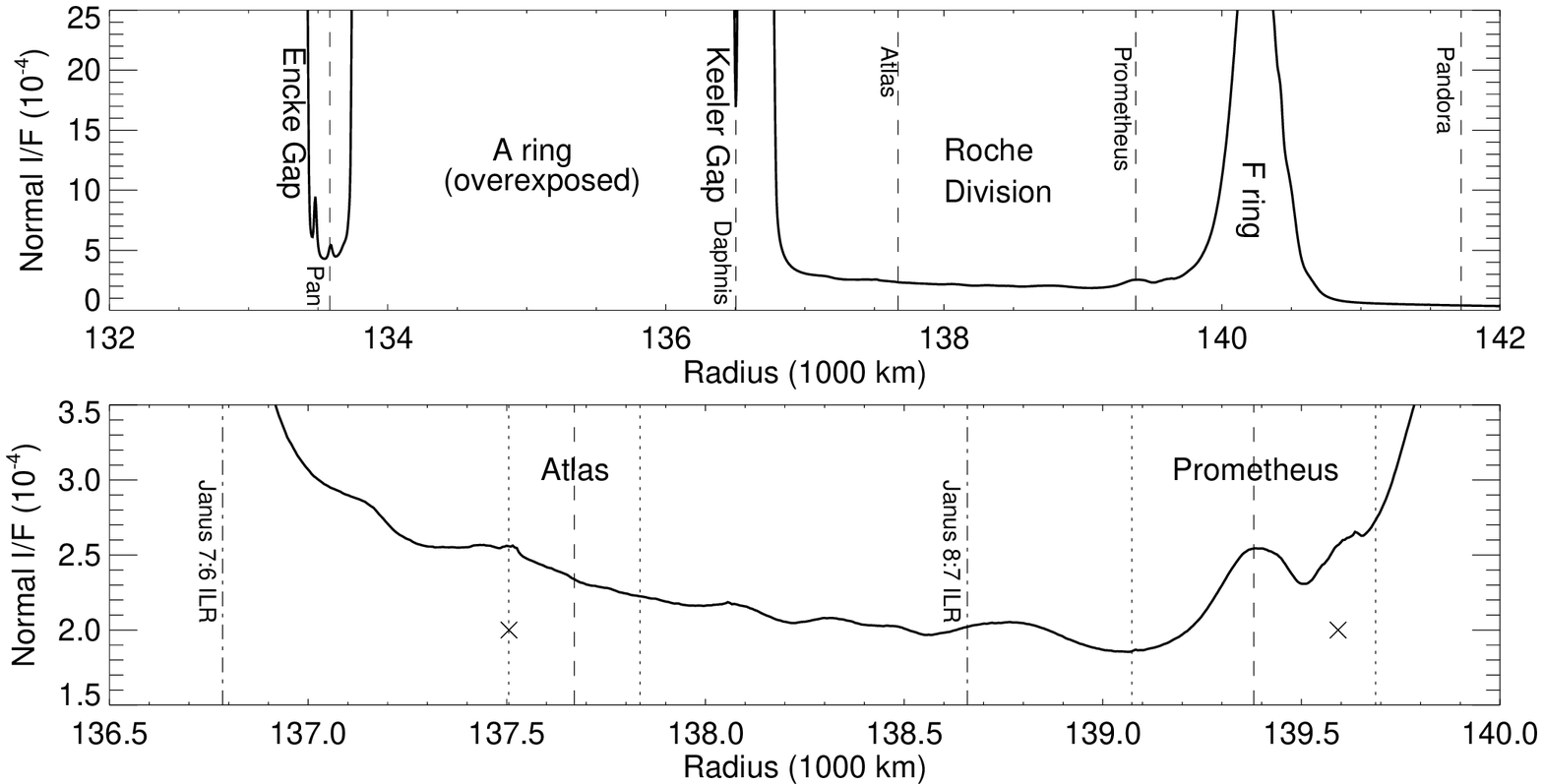}
\caption{On top is a labeled image (N1870374754 from Rev 268a), with arrows noting ring features and moons. The image has been stretched to highlight faint dusty structures. Note that the narrow Keeler Gap is not visible in this image because the stray light from the A ring causes it to appear artificially bright. Below are two radial scans of the image, showing the azimuthally averaged reflectance of the rings in Normal $I/F$ (see Section \ref{reduction}) vs. radial distance from Saturn's center. The lower plot is zoomed in on the Roche Division. Dashed lines mark the semi-major axes of each moon, while the dotted lines mark their pericenter and apocenter in the lower plot. Dot-dashed lines mark the locations of Janus/Epimetheus 1st order Lindblad resonances. The actual radial locations of Atlas and Prometheus at the time the image was taken are marked with X symbols. \label{rdradscan}}
\end{figure*}

The Roche Division is a sparsely populated region of Saturn's rings located between the outer edge of the A ring and the F ring. \citet{1983BAAS...15.1013B} first noted the presence of tenuous ring material between the A and F rings in a high phase angle \textit{Voyager 2} image of the region. At the time, it was referred to as the Pioneer Division, following \textit{Pioneer 11}'s discovery of the F ring \citep{1980Sci...207..434G}. \textit{Cassini} has since imaged the Roche Division many times, and Figure \ref{rdradscan} provides an overview of this region. The image shows several diffuse bands of dust populating almost the entire Roche Division. The Roche Division is also home to two of Saturn's innermost moons, Atlas and Prometheus, which stir up and provide a source of dust in their vicinity. The outermost dust band, at approximately $139,400$ km from Saturn's center, is typically the brightest feature in the region. \citet{2017Icar..281..322H} showed that this ringlet is in a co-orbital 1:1 resonance with Prometheus, but strangely precesses with the F ring. During \textit{Cassini's} Saturn orbit insertion, the \textit{Cassini} imaging team found two regions in the Roche Division with higher concentrations of dust than their surroundings \citep{2005Sci...307.1226P}. These apparent dusty ringlets were designated R/2004 S1 \citep{2004IAUC.8401....1P} orbiting near Atlas at $137$,$630$ km and R/2004 S2 \citep{2004IAUC.8432....1P} interior to Prometheus at $138$,$900$ km. In Figure \ref{rdradscan} we can see additional fainter dust bands between these two. \citet{2009Icar..202..260H} used image sequences taken in 2006 to determine that R/2004 S1 does not form a true closed ringlet, but a periodic structure composed of alternating diagonal bright and dark streaks. The structure is consistent with a dusty-ring's response to a strong resonance with a pattern speed near Saturn's rotation rate. The Roche Division is largely made up of tiny micron-sized particles that are sensitive to perturbations from non-gravitational forces. This led \citet{2009Icar..202..260H} to suspect that there is likely a connection with the magnetic field and/or radio emissions which were observed to be rotationally modulated at a rate commensurate with the Roche Division structure's pattern speed. 

Despite the perfect axisymmetry of Saturn's internally-generated magnetic field to within measurement accuracy \citep{2018Sci...362.5434D}, modulations near the $\sim10.5$ h planetary rotation period termed ``planetary period oscillations" (PPOs) are ubiquitously observed throughout the magnetosphere. Such modulations are observed in the magnetic field \citep{2000GeoRL..27.2785E,2010JGRA..115.4212A,2014JGRA..119.9847H,2018JGRA..123.3602B}, plasma properties and boundaries \citep{2007Sci...316..442G,2010JGRA..115.8209C,2011JGRA..11611205A,2017JGRA..122..393R,2017JGRA..122..280T}, energetic particle and associated energetic neutral atom fluxes \citep{1982GeoRL...9.1073C,2009GeoRL..3620103C,2011GeoRL..3816106C,2017JGRA..122..156C}, plasma waves \citep{2009GeoRL..3621108G,2010JGRA..11512258Y}, auroral ultraviolet and infrared emissions \citep{2010GeoRL..3715102N,2012JGRA..117.9228B,2018JGRA..123.8459B}, and auroral radio emissions \citep{1981GeoRL...8..253D,2011GeoRL..3821203G,2011JGRA..116.4212L,2013JGRA..118.4817L}. Analysis of the auroral radio emissions, specifically of Saturn kilometric radiation (SKR), provided the first evidence that two such modulation systems with slightly different periods are present \citep{2008JGRA..113.5222K,2009GeoRL..3616102G}, one associated with the northern polar ionosphere and the other with the southern, and that these periods vary slowly over Saturn's seasons by up to $\sim\pm1\%$ about a period of $\sim10.7$ h \citep{2000JGR...10513089G,2010GeoRL..3724101G}. Subsequently, the remotely sensed SKR emissions were used to derive the varying northern and southern PPO periods over the whole interval of the \textit{Cassini} science mission from January 2004 to end of mission in September 2017 \citep{2011pre7.conf...38L,2017pre8.conf..171L,2011pre7.conf...51G,2016JGRA..12111714Y,2018GeoRL..45.7297Y}. The periods and phases of the related magnetospheric magnetic perturbations have also been derived from \textit{Cassini} magnetometer (MAG) data from shortly after orbit insertion in mid-2004 to the end of mission \citep{2008JGRA..113.9205A,2012JGRA..117.4224A,2013JGRA..118.3243P,2016JGRA..121.9829P,2018JGRA..123.3859P}, apart from a few intervals when the perturbations due to one or other of the two PPO systems became too weak to be discerned. Comparison of the results derived independently from these data sets generally shows very good agreement \citep{2010JGRA..11512252A,2014JGRA..119.7380P,2016JGRA..121.9829P,2018JGRA..123.3859P}, as expected if both phenomena result from the same rotating magnetosphere-ionosphere current systems generated in the two polar ionospheres \citep{2012JGRA..117.4215J,2014JGRA..119.1563S}. Specifically both SKR and MAG results show that the southern PPO system was the stronger under Saturn southern summer conditions at the start of the \textit{Cassini} mission, and had a longer period $\sim10.8$ h ($800^{\circ}$day$^{-1}$) than that of the northern system $\sim10.6$ h ($815^{\circ}$day$^{-1}$). The two periods then converged towards a common value $\sim10.7$ h ($808^{\circ}$day$^{-1}$) around Saturn equinox in mid-2009, and after an interval of complex variable behavior, finally enduringly reversed in mid-2014 prior to northern solstice in mid-2017, with the stronger northern PPO system moving to a period of $\sim10.8$ h while the southern system period remained near $\sim10.7$ h. The magnetic oscillations of few nano-Tesla amplitude associated with the PPOs form a likely source of perturbations in the Roche Division. 

The inner Roche Division structures discovered by \citet{2009Icar..202..260H} were associated with the stronger longer period southern system in 2006. The Roche Division's microscopic ring particles are very effective at scattering incident sunlight in the forward direction and so are more easily observable in high-phase-angle (low scattering-angle) images. In this work we perform a comprehensive investigation of the high-phase-angle image sequences of the Roche Division obtained by the \textit{Cassini} Imaging Science Subsystem (ISS). Due to \textit{Cassini's} orbit geometry, the observations useful to this study are clumped into 3 distinct periods: 2006, 2012-2014, and 2016-2017. We find that the presence of periodic structures in the Roche Division is time variable. The structures are present in all 2006 observations as reported by \citet{2009Icar..202..260H}, but are not detected during any of the observation sequences from 2012-2014. The structures reappeared at some time before July of 2016, and are present in all later observations. This provides further evidence that the structures are somehow tied to Saturn's PPOs. 

We present the theoretical background of Lindblad resonances in planetary rings in Section 2, with particular emphasis on their framework in dusty rings, and our model of electromagnetic perturbations. We then report the observational data and data reduction procedures used in this analysis in Section 3. In Section 4, we detail the relevant pattern speeds and amplitudes from each epoch and make comparisons to the results of the magnetometer and radio and plasma wave science instruments. We discuss the significance of our detected pattern speeds and strengths in Section 5. Finally, in Section 6 we summarize our findings.

\section{Theoretical background}
In this work we use the model of Lindblad resonances in faint rings developed by \citet{2009Icar..202..260H}, to quantify the periodic structures in the Roche Division over the course of the \textit{Cassini} mission. This theory can account for ring structures produced by any periodic forcing that is commensurate with the ring particles' eccentric epicyclic motion and is not limited to classical Lindblad resonances associated with satellites. In dense rings, having sufficient mass and optical depth, Lindblad resonances produce density waves. Density waves have been studied extensively in Saturn's rings, particularly the A ring where many strong satellite resonances fall \citep{2007Icar..189...14T}. Strong Lindblad resonances in faint rings produce structures with a different appearance than those in dense rings. The sheets of dust in the Roche Division, with normal optical depth $\tau_n=(1-2)\times10^{-4}$ \citep{1998DPS....30.1705S}, lack the particle collision frequency and self-gravity necessary to sustain a normal spiral density wave. 
However, \citet{2009Icar..202..260H} found periodic azimuthal brightness variations in dusty rings at Lindblad resonances with Saturn's moon Mimas and with the rotation of Saturn's magnetic field. 

In Section \ref{lindblad} we discuss the general theory of Lindblad resonances. In Section \ref{3:4} we discuss the specific case of the dusty Roche Division particles responding to a Lindblad resonance. Finally in Section \ref{magmodel} we derive a simple model of an oscillating magnetic field component and determine the perturbed radial amplitude of the rings under particular conditions.

\subsection{Lindblad resonances \label{lindblad}}

For a ring particle in orbit around a central planet with a single satellite, the condition for a first-order Lindblad resonance is that the resonant argument:
\begin{equation}
\varphi=m(\lambda-\lambda_s)-(\lambda-\varpi) \label{resarg}
\end{equation}
is constant in time, $\frac{d\varphi}{dt}=0$. The inertial longitudes of the ring particle and satellite are $\lambda$ and $\lambda_s$ respectively and the ring particle's longitude of pericenter is $\varpi$. The integer $m$ is the azimuthal wavenumber of the resonance, and is positive (negative) for an inner (outer) Lindblad resonance, or ILR (OLR), where the ring particle is interior (exterior) to the satellite's orbit. This condition means that whenever the ring particle is in conjunction with the satellite, it will always be in the same phase of its eccentric orbit. For example, if $\lambda-\lambda_s=0$ (conjunction) occurs at the same time as $\lambda-\varpi=0$ (the ring particle is at periapse) then every time the ring particle approaches the satellite's longitude it will be back at its periapse. This allows the satellite's perturbations on the ring particle to build up and enhance the eccentricity of particles orbiting close to the resonance, as they always have close encounters with the perturbing moon in the same phase of their orbits. 

The expected forced radial amplitude, $ae_F$, on ring particles with semi-major axis, $a$, near a 1st order Lindblad resonance at $a_r$ is given by (\citet{1999ssd..book.....M} Equation 10.22)
\begin{equation}
ae_F\approx\frac{1.6a^2(M_s/M_p)}{3|a-a_r|}, \label{forced eccentricity}
\end{equation}
where $M_s$ and $M_p$ are the masses of the perturbing satellite and central planet respectively. In the absence of mutual interactions among ring particles the orbits of particles near a Lindblad resonance will follow a streamline path defined by their radius as a function of inertial longitude $\lambda$ and time $t$:
\begin{equation}
r(\lambda,t)=a+ae_F\cos{\phi}, \label{radii}
\end{equation}
where $\phi=m(\lambda-\Omega_p(t-t_0)-\delta_m)$ is the phase of the periodic structure. In a rotating frame the streamlines form $m$-lobed shapes with a specific orientation relative to the perturbing satellite, where $\delta_m$ is the inertial longitude of one of the structure's $m$-periapses ($a<a_r$) or $m$-apoapses ($a>a_r$) at epoch time $t_0$ and is roughly aligned with the longitude of the perturber. The structure rotates with a pattern speed $\Omega_p$ equivalent to the mean motion of the perturbing satellite. In dense rings, gravitational interactions among the ring particles cause neighboring streamlines located further and further from the resonance to shift in phase relative to the satellite, creating the appearance of an $m$-armed spiral density wave. Alternatively, a strong resonance near a ring's edge can form an $m$-lobed sinusoidal pattern as is the case for the outer edges of Saturn's B ring (Mimas 2:1 ILR) and A ring (Janus/Epimetheus 7:6 ILR) \citep{1984Icar...60...17P,2009AJ....138.1520S,2014Icar..227..152N}. Satellite resonances can also influence the entirety of narrow dense rings. This is the case for Uranus' $2$ km wide $\eta$ ring which has a $3$-lobed oscillation in its radius produced by the ring's proximity to Cressida's 3:2 ILR \citep{2017AJ....154..153C}. 

\subsection{3:4 Outer Lindblad Resonances in the Roche Division \label{3:4}}
The structures in the Roche Division have pattern speeds and symmetries consistent with $m=-3$ Outer Lindblad Resonances (OLR). No appropriate 1st-order Lindblad resonance with a Saturnian satellite falls within this region (for the Janus/Epimetheus 8:7 ILR see Section \ref{strength disc}), but \citet{2009Icar..202..260H} found that these structures were located coincident with the 3:4 resonance of the southern PPO system in 2006. 
By taking the time-derivative of Equation \ref{resarg} we obtain another expression that governs Lindblad resonances and normal modes:
\begin{equation}
m\Omega_p=(m-1)n+\dot{\varpi}=mn-\kappa \label{exppat}
\end{equation}
where $\frac{d\lambda_s}{dt}=\Omega_p$ is the pattern speed and mean motion of the perturber, $\frac{d\lambda}{dt}=n$ is the local mean motion, $\kappa$ is the local epicyclic frequency, and $\dot{\varpi}=n-\kappa$ is the local apsidal precession rate. Note that $n$ and $\kappa$ are the angular and epicyclic frequencies, corrected for the effects of an oblate central planet \citep{1982AJ.....87.1051L}. The 3-lobed structures generated by such a resonance that could occur between the outer edge of the A ring (mean radius of $136,770$ km) and the orbit of Prometheus ($a=139,380$ km) are expected to have pattern speeds of $804.64^{\circ}$day$^{-1}$ to $782.11^{\circ}$day$^{-1}$ respectively, or roughly $4/3$ the mean motion of ring particles orbiting in the Roche Division ($n=604.22^{\circ}$day$^{-1}$ to $587.27^{\circ}$day$^{-1}$). Thus, in image-movies that span about one local orbital period we see $4/3$ of the structure pass through the imaged longitudinal region. Note that if we construct mosaics in a co-rotating longitude system using the expected pattern speed of $\sim800^{\circ}$day$^{-1}$ we see the full 3-fold symmetric pattern. 

Individual dusty ring particles will also respond to gravitational perturbations near a strong Lindblad resonance. However, in the case of dusty rings the particle streamlines appear to be ``damped" to form a region of canted brightness variations straddling the resonance. \citet{2009Icar..202..260H} created a model to explain the observed behavior of dusty ring structures in Saturn's Roche Division, D ring, and G ring by introducing a dissipative term to the equations of motion. This presumed dissipation causes a phase shift of ring particle streamlines that depends on their radial distance from the resonance and results in the observed canted brightness variations (see Figure \ref{example mosaic} below). 

For a low optical depth region, like the Roche Division, the fractional brightness variations are a small percentage of the overall brightness and are effectively equivalent to the local fractional surface density variations. In their model of Lindblad resonances in faint rings, \citet{2009Icar..202..260H} found that the fractional surface density variations $\frac{\delta\rho}{\rho}$ near a resonance in a dusty ring are given by:
\begin{equation}
\frac{\delta\rho}{\rho}(r,\phi)=\frac{\beta}{[(\delta r)^2+L^2]^2}[((\delta r)^2-L^2)\cos{\phi}+2\delta rL\sin{\phi}], \label{particle density}
\end{equation}
where $\delta r$ is the radial distance from the exact resonance, $\beta$ is a measure of the resonance strength, and $L$ is a damping length. The amplitude of the brightness variations
\begin{equation}
A(r)=\frac{\beta}{(\delta r)^2+ L^2} \label{amplitude}
\end{equation}
due to the density variations reaches a peak at the exact resonance. Additionally, this amplitude of the brightness variations is proportional to the strength of the resonance. A full derivation of the relevant equations from \citet{2009Icar..202..260H} is reiterated in Appendix \ref{derivations} with some corrections. 

\citet{2009Icar..202..260H} found that the G ring resonant structure straddles the Mimas 8:7 ILR, such that its maximum brightness variations do in fact peak at the exact resonance\footnote{However, the reported amplitude of the variations is actually 2.5 times larger than one would expect, so the estimated value of $\beta=\frac{1.6a^2}{3}\frac{M_s}{M_p}$ overestimates the mass of Mimas by a factor of 2.5, see Section \ref{strength disc} below.}. Thus, we can determine the exact resonance locations even when we do not know the exact angular frequency of the perturber and also approximate the strength of the resonance in terms of a satellite mass when we do not know the mass of the perturber. Alternatively, if these structures are generated by Saturn's magnetic field, the perturbing force is more likely the Lorentz force, and so we can estimate the amplitude of the magnetic-field variations. 

\subsection{Simple magnetic field perturbation model \label{magmodel}}
As we have explained above, it is most likely that the dust in the Roche Division is perturbed at the 3:4 OLR with the rotational modulation of Saturn's magnetosphere. The case for resonances with the rotation of planetary magnetic fields, or Lorentz resonances, has been explored to explain the structure of Jupiter's faint rings \citep{1985Natur.316..115B}. In fact, the theory has been derived extensively in the works of \citet{1992Icar...96...65S} and \citet{1994Icar..109..221H}. We summarize those derivations in Appendix \ref{lorentz}, but note that in order to approximate the strength of the 3:4 Lorentz resonance in question, one needs to know the planet's non-axisymmetric magnetic field coefficients. However, current models of Saturn's magnetic field \citep{2018Sci...362.5434D} show that the planet has negligible non-axisymmetric components, so the relevant magnetic field perturbations are probably external to the planet. Thus, we improvise by instituting a simple sinusoidal perturbation in the magnetic field that resembles the observed planetary period oscillations measured by \textit{Cassini's} magnetometer.

We begin with the Lorentz Force $\mathbf{F_L}=q_g\mathbf{v_{rel}}\times\mathbf{B}$ on a charged dust grain $q_g$, which arises from the rotation of Saturn's magnetic field $\mathbf{B}$. Here $\mathbf{v_{rel}}$ is a ring particle's orbital velocity relative to the angular frequency of Saturn's PPO rotation $\Omega_S$ given by
\begin{equation}
\mathbf{v_{rel}}=\mathbf{v}-(\bm{\Omega_s}\times\mathbf{r})
\end{equation}
where $\mathbf{r}$ and $\mathbf{v}$ are a ring particle's radial position and Keplerian orbital velocity in the ring plane. We are considering low eccentricity rings around Saturn, whose magnetic pole is effectively aligned with its rotation axis \citep{2018Sci...362.5434D}, so we can simplify this expression to
\begin{equation}
\mathbf{v_{rel}}=a(n-\Omega_S)\bm{\hat{\phi}}.
\end{equation}
Our Lorentz force is then given by
\begin{equation}
\mathbf{F_L}=-q_ga(n-\Omega_S)\left[B_{\theta}\mathbf{\hat{r}}-B_r\bm{\hat{\theta}}\right]
\end{equation}
where the radius $r$, polar angle $\theta$, and azimuthal angle $\phi$ define the usual spherical coordinates relative to the rotating planet. We mainly care about the small PPO magnetic perturbation on top of the static $B_{\theta}$ component of the magnetic field because it will produce an oscillating force in the radial direction, as needed to create a Lindblad resonance. While the fundamental PPO magnetospheric signal will go like $\cos{\left(\lambda-\Omega_St\right)}$, deviations in the shape of the magnetic field perturbation from a pure sinusoid and nonlinearities in the expression for the perturbing force will give rise to harmonics of the form $\cos{\left[j(\lambda-\Omega_St)\right]}$ for various integer values of $j$. We consider a simple perturbation $B_{\theta0}\cos{\left[j(\lambda-\Omega_St)\right]}$ where $B_{\theta0}$ can be compared to measured MAG oscillations modulated by the rotation periods associated with the northern and southern PPO systems \citep{2018JGRA..123.3859P}. 

The perturbation produces a radial force given by
\begin{equation}
F_r=-q_ga(n-\Omega_S)B_{\theta0}\cos{\left[j(\lambda-\Omega_St)\right]}.
\end{equation}
This perturbing force can now be inserted into the perturbation equations of a ring particle's eccentricity and longitude of pericenter, which for objects on nearly circular orbits are \citep{1976AmJPh..44..944B,1977AmJPh..45.1230B}
\begin{equation}
\frac{de}{dt}=n\left[\frac{F_r}{F_G}\sin{f}+2\frac{F_{\lambda}}{F_G}\cos{f}\right]
\end{equation}
and
\begin{equation}
\frac{d\varpi}{dt}=\frac{n}{e}\left[-\frac{F_r}{F_G}\cos{f}+2\frac{F_{\lambda}}{F_G}\sin{f}\right]
\end{equation}
where the ring particle's true anomaly $f=\lambda-\varpi$ and the central gravitational force on a dust grain of mass $m_g$ due to Saturn is $F_G=\frac{GM_Sm_g}{a^2}$. After plugging in our $F_r$ (assuming the longitudinal force $F_{\lambda}=0$) and applying the appropriate trigonometric product and sum identities these become:
\begin{widetext}
\begin{equation}
\frac{de}{dt}=\frac{-nq_ga^3(n-\Omega_S)B_{\theta0}}{GM_Sm_g}\frac{1}{2}\{\sin{\left[j(\lambda-\Omega_St)+(\lambda-\varpi)\right]}-\sin{\left[j(\lambda-\Omega_St)-(\lambda-\varpi)\right]}\}
\end{equation}
and
\begin{equation}
\frac{d\varpi}{dt}=\frac{nq_ga^3(n-\Omega_S)B_{\theta0}}{eGM_Sm_g}\frac{1}{2}\{\cos{\left[j(\lambda-\Omega_St)+(\lambda-\varpi)\right]}+\cos{\left[j(\lambda-\Omega_St)-(\lambda-\varpi)\right]}\}.
\end{equation}
\end{widetext}
The perturbations on the ring particles are small, so we can assume that $\lambda \approx nt$ and $\varpi \approx \dot{\varpi_0}t$. Hence, for particles near a particular resonance with a specified value of $m$, these expressions will nearly all average to zero. The only case where this does not happen is when $j=m$, in which case the first term will average to zero and the second can be written with $\varphi=m(\lambda-\Omega_St)-(\lambda-\varpi)$. Hence the time-average perturbation questions become:
\begin{equation}
\left\langle\frac{de}{dt}\right\rangle=\frac{nq_ga^3(n-\Omega_S)B_{\theta0}}{2GM_Sm_g}\sin{\varphi}
\end{equation}
and
\begin{equation}
\left\langle\frac{d\varpi}{dt}\right\rangle=\frac{nq_ga^3(n-\Omega_S)B_{\theta0}}{2eGM_Sm_g}\cos{\varphi}+\dot{\varpi_0}.
\end{equation}
Following the logic used to derive Equation \ref{gforcede} we find the forced eccentricity is \begin{equation}
a_re_f=\frac{1}{3}\frac{q_g}{m_g}\frac{a_r^5}{\delta a}\frac{n/\Omega_S-1}{m-1}\frac{\Omega_S}{GM_S}B_{\theta0}
\end{equation}
where our resonance strength is given by
\begin{equation}
\beta=\frac{1}{3}\frac{q_g}{m_g}a_r^5\frac{n/\Omega_S-1}{m-1}\frac{\Omega_S}{GM_S}B_{\theta0}.
\end{equation}

Finally for the case of the Roche Division, where $m=-3$ and $\frac{n}{\Omega_S} \approx \frac{3}{4}$ we find:
\begin{equation}
\beta=\frac{1}{48}\frac{q_g}{m_g}a_r^5\frac{\Omega_S}{GM_S}B_{\theta0} \label{bamp}
\end{equation}
which can be used to estimate the amplitude of magnetic field oscillations needed to produce the observed structures in the rings.
\pagebreak

\startlongtable
\begin{longrotatetable}
\begin{deluxetable}{cccccccccc}
\tablecaption{Imaging sequences used in this study \label{imaging data}}
\tablewidth{0pt}
\tablehead{
\colhead{Image} &
\colhead{Observation ID} &
\colhead{Files} &
\colhead{\#} &
\colhead{Observation} &
\colhead{Duration} &
\colhead{Exposure} &
\colhead{Phase} &
\colhead{Emission} &
\colhead{Inertial} \\
\colhead{movie} & & & & \colhead{date} & \colhead{(hours)} & \colhead{(seconds)} & \colhead{angle} & \colhead{angle} & \colhead{longitude} \\
\colhead{ID} & & & & & & & \colhead{($^{\circ}$)} & \colhead{($^{\circ}$)} & \colhead{($^{\circ}$)} \\
}
\startdata
029ax &     ISS{\_}029RF{\_}FMOVIE001{\_}VIMS & N1538168640-N1538193384 &  24 & 2006 SEP 28 &  6.9 & 0.18 & 158.4-160.2 &  57.5- 58.3 & 265.6-267.7 \\
029ay &     ISS{\_}029RF{\_}FMOVIE001{\_}VIMS & N1538269441-N1538275217 &   6 & 2006 SEP 30 &  1.6 & 0.18 & 158.9-159.3 &  60.6- 60.8 &  83.4- 83.9 \\
029bx &     ISS{\_}029RF{\_}FMOVIE001{\_}VIMS & N1538169174-N1538217594 &  46 & 2006 SEP 28 & 13.4 & 0.68 & 158.5-161.7 &  57.5- 59.1 & 265.7-269.5 \\
029by &     ISS{\_}029RF{\_}FMOVIE001{\_}VIMS & N1538270015-N1538300071 &  27 & 2006 SEP 30 &  8.3 & 0.68 & 158.9-161.0 &  60.6- 61.5 &  83.5- 85.7 \\
029cx &     ISS{\_}029RF{\_}FMOVIE001{\_}VIMS & N1538194460-N1538218132 &  23 & 2006 SEP 29 &  6.6 & 0.15 & 160.3-161.8 &  58.3- 59.1 & 267.8-269.6 \\
029cy &     ISS{\_}029RF{\_}FMOVIE001{\_}VIMS & N1538276373-N1538299493 &  21 & 2006 SEP 30 &  6.4 & 0.15 & 159.4-160.9 &  60.8- 61.5 &  84.0- 85.7 \\
 030a &     ISS{\_}030RF{\_}FMOVIE001{\_}VIMS & N1539655570-N1539660972 &   4 & 2006 OCT 16 &  1.5 & 0.46 & 150.6-151.0 &  48.7- 48.9 & 265.0-265.5 \\
 030b &     ISS{\_}030RF{\_}FMOVIE001{\_}VIMS & N1539656467-N1539683497 &  16 & 2006 OCT 16 &  7.5 & 1.00 & 150.6-152.4 &  48.7- 49.7 & 265.1-267.5 \\
 030c &     ISS{\_}030RF{\_}FMOVIE001{\_}VIMS & N1539662774-N1539682596 &  12 & 2006 OCT 16 &  5.5 & 0.38 & 151.1-152.4 &  48.9- 49.6 & 265.7-267.4 \\
 031a &     ISS{\_}031RF{\_}FMOVIE001{\_}VIMS & N1541012989-N1541025000 &  28 & 2006 OCT 31 &  3.3 & 0.26 & 156.3-157.4 &  51.9- 52.7 &  95.0- 96.2 \\
 031b &     ISS{\_}031RF{\_}FMOVIE001{\_}VIMS & N1541025445-N1541045470 &  46 & 2006 OCT 31 &  5.6 & 0.22 & 157.4-159.0 &  52.7- 54.1 &  96.2- 98.2 \\
031cx &     ISS{\_}031RF{\_}FMOVIE001{\_}VIMS & N1541045915-N1541062380 &  38 & 2006 NOV 01 &  4.6 & 0.18 & 159.1-160.3 &  54.1- 55.2 &  98.2- 99.7 \\
031cy &     ISS{\_}031RF{\_}FMOVIE001{\_}VIMS & N1541110189-N1541119755 &  12 & 2006 NOV 01 &  2.7 & 0.18 & 159.9-160.0 &  58.3- 58.9 & 294.3-294.9 \\
 032a &     ISS{\_}032RF{\_}FMOVIE001{\_}VIMS & N1542047155-N1542063616 &  38 & 2006 NOV 12 &  4.6 & 0.26 & 155.9-157.3 &  51.8- 52.9 &  94.7- 96.4 \\
 032b &     ISS{\_}032RF{\_}FMOVIE001{\_}VIMS & N1542064061-N1542084086 &  46 & 2006 NOV 12 &  5.6 & 0.22 & 157.4-159.0 &  52.9- 54.3 &  96.4- 98.3 \\
 032c &     ISS{\_}032RF{\_}FMOVIE001{\_}VIMS & N1542084531-N1542096546 &  28 & 2006 NOV 13 &  3.3 & 0.18 & 159.1-160.0 &  54.3- 55.1 &  98.4- 99.4 \\
 032d &     ISS{\_}032RF{\_}FMOVIE001{\_}VIMS & N1542149816-N1542156952 &  18 & 2006 NOV 13 &  2.0 & 0.12 & 163.5-163.9 &  58.5- 59.0 & 103.7-104.2 \\
  033 &     ISS{\_}033RF{\_}FMOVIE001{\_}VIMS & N1543166702-N1543216891 &  99 & 2006 NOV 25 & 13.9 & 0.56 & 159.8-160.7 &  57.6- 60.7 & 293.5-296.5 \\
  036 &     ISS{\_}036RF{\_}FMOVIE001{\_}VIMS & N1545557060-N1545613256 & 127 & 2006 DEC 23 & 15.6 & 0.56 & 158.6-160.8 &  54.2- 57.2 & 285.9-289.7 \\
 039a &     ISS{\_}039RF{\_}FMOVIE002{\_}VIMS & N1549801218-N1549851279 & 123 & 2007 FEB 10 & 13.9 & 2.60 & 125.5-131.1 &  31.4- 33.1 & 243.2-254.8 \\
 039b &     ISS{\_}039RF{\_}FMOVIE002{\_}VIMS & N1549901779-N1549911779 &  26 & 2007 FEB 11 &  2.8 & 3.20 & 136.6-137.7 &  35.8- 36.5 & 265.1-267.0 \\
\hline
 173ax &    ISS{\_}173RF{\_}FMOVIE001{\_}PRIME & N1729024626-N1729053296 &  55 & 2012 OCT 15 &  8.0 & 1.80 & 137.5-142.1 & 123.9-124.2 & 138.4-112.4 \\
 173by &    ISS{\_}173RF{\_}FMOVIE001{\_}PRIME & N1729053606-N1729082276 &  55 & 2012 OCT 16 &  8.0 & 1.20 & 153.4-160.8 & 123.3-118.4 & 334.3-308.4 \\
 174x & ISS{\_}174RF{\_}FRSTRCHAN001{\_}PRIME & N1731106419-N1731132308 &  63 & 2012 NOV 08 &  7.2 & 1.20 & 150.4-156.0 & 123.6-121.2 & 318.5-318.5 \\
 174y & ISS{\_}174RF{\_}FRSTRCHAN001{\_}PRIME & N1731132699-N1731158588 &  64 & 2012 NOV 09 &  7.2 & 1.20 & 143.9-149.5 & 121.6-119.3 & 135.0-135.1 \\
 185a &  ISS{\_}185RI{\_}ROCHEMOV001{\_}PRIME & N1743541318-N1743581038 &  31 & 2013 APR 01 & 11.0 & 2.00 & 139.4-162.7 & 134.5-116.1 & 138.9-159.4 \\
  196 &    ISS{\_}196RF{\_}FMOVIE006{\_}PRIME & N1756377239-N1756428571 & 125 & 2013 AUG 28 & 14.3 & 1.20 & 145.3-150.1 & 132.7-129.5 & 301.1-304.6 \\
  201 &     ISS{\_}201RF{\_}FMOVIE001{\_}VIMS & N1770315948-N1770367142 & 160 & 2014 FEB 05 & 14.2 & 1.20 & 125.8-142.8 & 137.6-131.0 & 305.0-327.3 \\
 203x &    ISS{\_}203RF{\_}FMOVIE001{\_}PRIME & N1776077815-N1776106170 &  70 & 2014 APR 13 &  7.9 & 1.20 & 139.8-136.5 &  74.5- 72.1 & 203.1-202.8 \\
 203y &    ISS{\_}203RF{\_}FMOVIE001{\_}PRIME & N1776106495-N1776134850 &  70 & 2014 APR 13 &  7.9 & 1.20 & 132.0-129.0 &  72.2- 70.3 &  37.2- 37.2 \\
 208x &    ISS{\_}208RF{\_}FMOVIE001{\_}PRIME & N1789913768-N1789932276 &  90 & 2014 SEP 20 &  5.1 & 1.20 & 131.9-142.4 & 132.6-126.5 &  51.6- 51.2 \\
  209x &    ISS{\_}209RF{\_}FMOVIE001{\_}PRIME & N1793068408-N1793089959 &  46 & 2014 OCT 27 &  6.0 & 1.20 & 134.2-132.8 &  72.0- 71.2 & 218.6-218.6 \\
 209y &    ISS{\_}209RF{\_}FMOVIE001{\_}PRIME & N1793090708-N1793110822 &  43 & 2014 OCT 27 &  5.6 & 1.20 & 129.1-127.9 &  71.2- 70.5 &  44.1- 44.2 \\
\hline
  237 & ISS{\_}237RI{\_}EGAPMOVMP001{\_}PRIME & N1846112392-N1846137205 &  28 & 2016 JUL 01 &  6.9 & 0.32 & 135.3-130.5 &  83.8- 80.7 & 243.8-247.7 \\
  240 &    ISS{\_}240RF{\_}FMOVIE002{\_}PRIME & N1850609421-N1850636617 &  50 & 2016 AUG 23 &  7.6 & 1.50 & 134.6-130.2 &  81.0- 77.5 & 245.4-246.1 \\
  241 &  ISS{\_}241RI{\_}FNTHPMOV001{\_}PRIME & N1851525687-N1851564003 &  32 & 2016 SEP 02 & 10.6 & 2.00 & 159.7-149.9 & 100.2- 92.7 & 228.8-235.1 \\
  265 &    ISS{\_}265RF{\_}FMOVIE001{\_}PRIME & N1867972213-N1867991971 &  30 & 2017 MAR 12 &  5.5 & 1.20 & 134.3-132.0 &  74.8- 72.6 & 233.6-233.6 \\
 268a & ISS{\_}268RI{\_}HPMONITOR002{\_}PRIME & N1870352908-N1870396600 &  23 & 2017 APR 08 & 12.1 & 2.00 & 145.6-140.7 &  84.5- 80.1 & 238.3-241.6 \\
  272 &    ISS{\_}272RF{\_}FMOVIE001{\_}PRIME & N1872298730-N1872325959 &  74 & 2017 MAY 01 &  7.6 & 1.00 & 136.5-133.7 &  75.3- 72.6 & 237.6-238.2 \\
 274x &    ISS{\_}274RF{\_}FMOVIE001{\_}PRIME & N1873378488-N1873395652 &  75 & 2017 MAY 13 &  4.8 & 1.00 & 139.9-138.5 &  78.0- 76.6 & 233.8-234.3 \\
 274y &    ISS{\_}274RF{\_}FMOVIE001{\_}PRIME & N1873415748-N1873456613 & 172 & 2017 MAY 14 & 11.4 & 1.00 & 130.3-124.4 &  74.3- 69.8 &  83.8- 84.2 \\
  281 & ISS{\_}281RI{\_}HPMONITOR001{\_}PRIME & N1877745556-N1877791066 &  38 & 2017 JUL 03 & 12.6 & 2.00 & 147.3-144.3 &  84.1- 81.0 & 221.2-221.1 \\
  287 & ISS{\_}287RI{\_}HPMONITOR001{\_}PRIME & N1880562754-N1880589544 &  31 & 2017 AUG 04 &  7.4 & 2.00 & 145.0-143.0 &  82.1- 80.3 &  54.7- 56.3 \\
  289 &    ISS{\_}289RF{\_}FMOVIE001{\_}PRIME & N1881776562-N1881839798 & 156 & 2017 AUG 18 & 17.6 & 1.00 & 137.8-131.5 &  75.0- 68.4 & 226.4-228.1 \\
  292 &    ISS{\_}292RF{\_}FMOVIE001{\_}PRIME & N1883393812-N1883445888 &  63 & 2017 SEP 06 & 14.5 & 0.68 & 142.2-138.2 &  78.9- 75.1 &  42.5- 42.8 \\
\enddata
\tablecomments{Inertial longitudes are measured relative to the ascending node of the rings on J2000. Image movie IDs are a combination of Rev number, exposure duration (a, b, or c), and ring ansa (x or y). Observations such as 029ay and 029cy are too short to be useful on their own, but can be combined with each other and 029by (see Figure \ref{rev0292}).}
\end{deluxetable}
\end{longrotatetable}

\section{Observational data}
We conducted a comprehensive survey of the available Roche Division images from the \textit{Cassini} cameras. Section \ref{data} summarizes the observations used in this study, while Section \ref{reduction} describes how these data were processed to obtain maps of the fractional brightness variations that can be compared with the above model predictions.
\subsection{Imaging data \label{data}}
This study examines high-phase-angle image sequences of the Roche Division obtained over the course of the entire \textit{Cassini} mission by the Imaging Science Subsystem (ISS). For consistency, we only consider images obtained by the Narrow Angle Camera (NAC) using the clear filters, because they are the most common and have the highest signal-to-noise. Every image is calibrated using the standard CISSCAL calibration routines \citep{2004SSRv..115..363P,2010P&amp;SS...58.1475W} to remove backgrounds, flatfield, and convert the raw data numbers into $I/F$, a standard measure of reflectance. $I$ is the intensity of radiation scattered by the ring particles and $\pi F$ is the solar flux at Saturn, so $I/F$ is a unitless quantity that equals unity for a perfect Lambert surface viewed at normal incidence. We geometrically navigate the images using appropriate SPICE kernels \citep{1996P&amp;SS...44...65A} and by matching background point sources or streaks to stars in the Tycho and UCAC catalogs.

The images chosen for this analysis belong to movie sequences. These are image sequences taken of a common longitudinal region of the rings over the course of a large fraction of the local ring region's orbital period. This enables us to watch ring material move through the region over time, so we can detect radial and azimuthal brightness variations in the rings. We've found that the structures we are interested in are only detectable in images taken at phase angles $\alpha$ above around $130^{\circ}$. 

Table \ref{imaging data} shows a breakdown of the complete dataset used in this study. \textit{Cassini's} observations of the Roche Division were conducted during three distinct epochs, separated by the horizontal lines in the table. 
Throughout this work we identify the individual image sequences by their Saturn orbit numbers (or Rev number), exposure durations (a,b,c), and ring ansa (x or y) when necessary. This is not to be confused with the mission assigned Observation ID used to identify the images within archives, such as the PDS rings node. Table \ref{imaging data} also contains the ring ansa longitudes and lighting geometries spanned by the image sequences. Any individual images containing  substantial defects or insufficient background stars for proper pointing have been removed from our in-depth analysis, but in some cases are useful merely for identifying the presence of structures (for instance Rev 289 has defects preventing proper analysis, but the presence of structures is clear). We also leave out observations that are less than 3.6 hours long, which are too short to reliably identify the resonant structures.

\subsection{Data reduction \label{reduction}}

\begin{figure*}[ht]
\includegraphics[width=\linewidth]{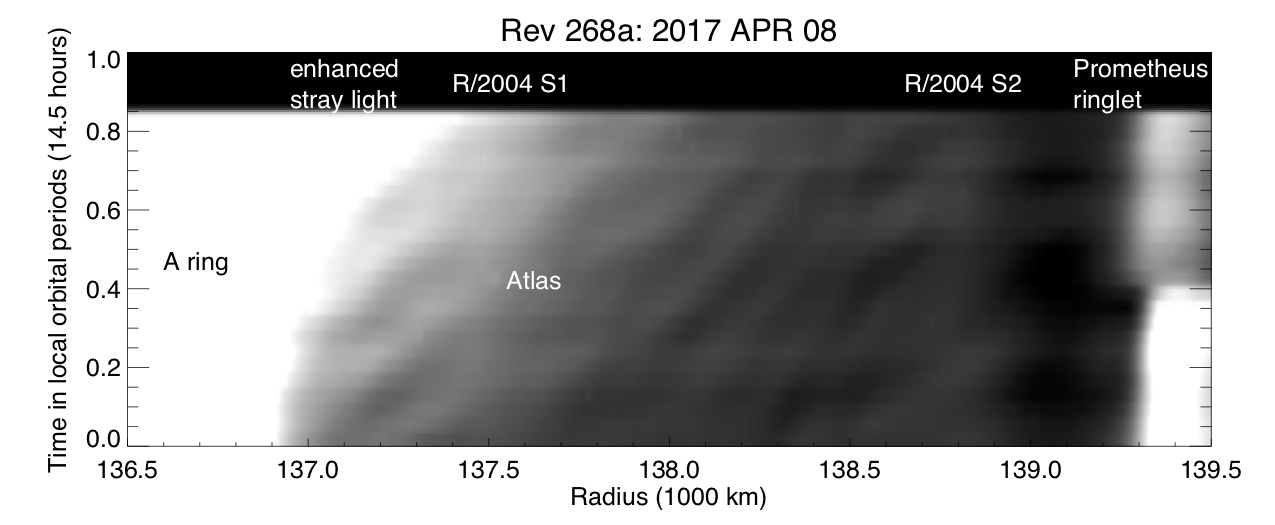}
\caption{This map of the Roche Division is made up of average brightness radial profiles of individual images, at an inertial longitude of $\sim240^{\circ}$, stacked from bottom to top sequentially in time and shown in local orbital periods. The bright A ring edge is at the left of the map and the Prometheus ringlet is at the right, with a radial and brightness asymmetry clearly displayed on either side of Prometheus which passed by this longitude about halfway through the sequence (Atlas passed by at the same time and its faint brightness signature is labeled). The Roche Division is brighter in the upper left corner of the map because the ring opening angle is slowly increasing, resulting in a brighter A ring edge. There are multiple regions of diagonal brightness variations between $137,000$ and $139,000$ km.   \label{example mosaic}}
\end{figure*}

Each image of the Roche Division covers a relatively small range of longitude compared to the scale of the azimuthal brightness variations. Thus, we reduce each image into a high signal-to-noise profile of ring brightness as a function of radius by averaging the image's brightness data over all longitudes spanned by the image. 

The lighting geometry in each image can vary significantly both within an individual sequence and between observations taken at different epochs. We can compare the radial profiles of images with different emission angles $e$ or ring opening angles $B=90^{\circ}-e$ (not to be confused with the magnetic field) by converting the observed $I/F$ reflectance into ``normal $I/F$'' or $\mu I/F$, where $\mu=|\sin{B}|$. Ideally, for a low optical depth region like the Roche Division, the normal $I/F$ should be independent of incidence and emission angle. However, even after this correction several obstacles remain when trying to compare images from multiple sequences or even within individual sequences. 

While several observations contain images with nearly identical phase angles, others have a gradual change in phase angle, sometimes by more than $10^{\circ}$ over the course of the observation. The range of phase angles from all images included in the study is $125^{\circ}-163^{\circ}$, making direct comparison of the normal $I/F$ profiles problematic. The general trend for dusty rings is that they appear brighter at higher phase angles, which can be described in terms of a phase function. However, in this case further difficulties arise because of the background signals from the A and F rings. 

There are drastic variations in stray light from the A ring among the different data sets because they include observations of both the lit and unlit side of the rings. Images of the lit side of Saturn's rings capture a bright A ring whose brightness can leak into the Roche Division due to the stray light and extended PSF of the camera \citep{2010P&amp;SS...58.1475W}. Raw radial scans of these images exhibit a bright inner Roche Division that generally decreases in brightness with increasing radii until approaching the inner edge of the Prometheus ringlet and the F ring (see Figures \ref{rdradscan} and \ref{example mosaic}). Conversely, radial scans of the unlit side of the rings, which include a darker A ring, often reveal a much darker inner Roche Division with a general brightness enhancement contaminating the outer Roche Division as one approaches the F ring. Additionally, calibrated images with different exposure durations with otherwise nearly identical observation geometries can have constant instrumental offsets that produce slightly different reflectance levels in images of the Roche Division that are relevant here because the rings themselves are so faint.

\begin{figure*}[ht]
\includegraphics[width=\linewidth]{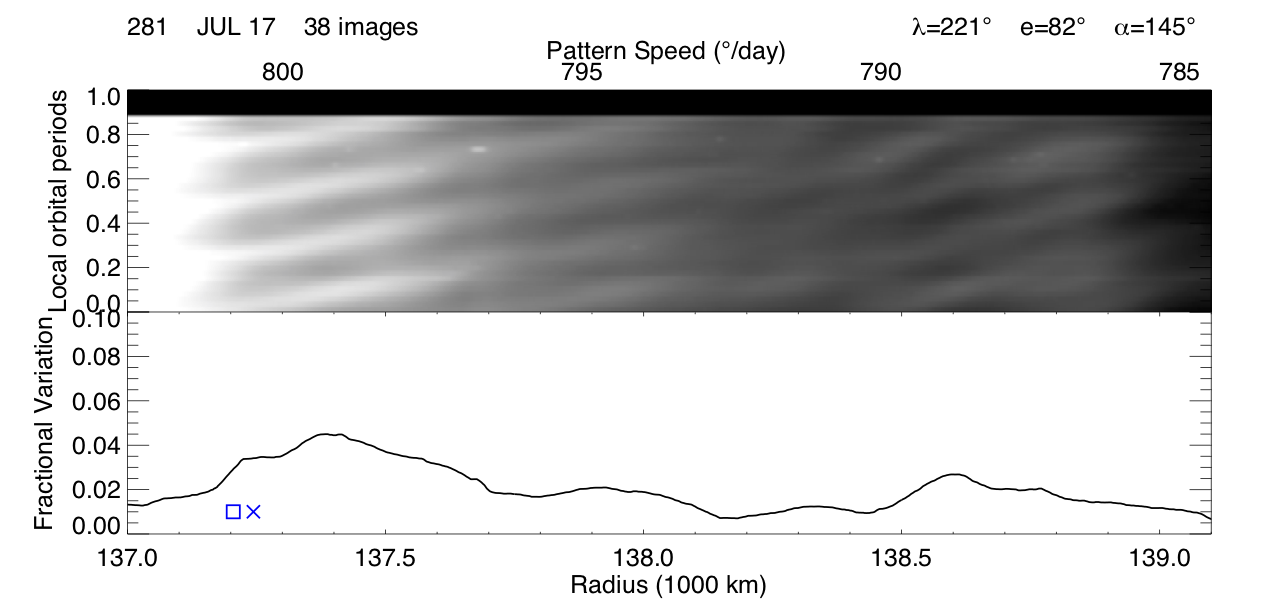}
\caption{Above is a corrected brightness map of the Roche Division from Rev 281 similar to Figure \ref{example mosaic}, atop the fractional variation in the rings brightness at each radius and pattern speed. The SKR (X) and MAG (square) rate/radius are marked in blue for the northern hemisphere (red for the southern hemisphere at early times). The peak of $4.5\%$ occurs at $799.2$ $^{\circ}$day$^{-1}$ or $137,385$ km. When the resonant structures are detected, those with the largest fractional brightness variations consistently occur in this inner region of the Roche Division near the magnetospheric periodicities. \label{example fracamp}}
\end{figure*}

These issues are best illustrated by considering a single data set. Figure \ref{example mosaic} is an example map of the Roche Division, created using the radial scans from Rev 268a images. The Rev 268a image-movie monitored the Roche Division for 12.1 hours (about 0.86 local orbital periods). The radial profile of the first image is on the bottom, with the following images' radial scans shown in the map sequentially to the top. In this particular movie, the phase angle of each image decreases over the course of the observation from bottom ($\alpha=145^{\circ}$) to top ($\alpha=140^{\circ}$), which should normally cause a decrease in the observed reflectance due to less optimal forward scattering. In reality, brightness increases from bottom to top in the left side of the image because of increasing amounts of stray light from the A ring. The ring plane opening angles in this observation are all $<10^{\circ}$ and vary by $4^{\circ}$ over the course of the observation. At these small opening angles the A-ring's brightness depends strongly on the viewing geometry. In general, images with larger opening angles suffer from increased contamination of the bright A ring on the lit side of the rings after converting the profiles to Normal $I/F$. This means that as the ring opens up (smaller $e$, larger $B$) the Roche Division reflectance is artificially enhanced. 
This increase is especially apparent near the A ring in the upper left corner of the mosaic. Whenever the ring opening angle and emission angle vary significantly, there is a similar alteration in ring brightness near the F ring in observations of the unlit side of the rings. We should also note that the Prometheus ringlet, located in the far right of this figure, has its own individual asymmetry in reflectance and radius on either side of Prometheus. This region is far enough from the A ring to be largely unaffected by the changing ring opening angle. However, the decrease in phase angle over the course of the observation serves to enhance this characteristic asymmetry observed in the Prometheus ringlet reflectance.

Despite these distracting trends in reflectance due to the lighting geometry, in Figure \ref{example mosaic} we see that the bands of dust in individual images (e.g. Figure \ref{rdradscan}) change in radius over time (we recommend viewing the mosaic figures in the electronic copy). In fact, the ring brightness in each radial region is clearly periodic, appearing to increase and decrease four times over the course of an orbital period. In this observation there appear to be at least three distinct canted azimuthal brightness variations centered at approximately $137,300$ km, $138,200$ km, and $138,600$ km. Each of these azimuthal brightness variations look like the characteristic response of dusty rings to a strong resonance described in Section \ref{3:4} above.

Besides these patterns of canted bands, there are also structures in this region that might be tied to various moons. More continuous dust bands may occupy the narrow regions around $137,500$ and $138,800$ km, possibly corresponding to R/2004 S1\&2 and having some relation to the orbit of Atlas and the position of the Janus/Epimetheus 8:7 ILR, respectively. Upon further stretching of the image in Figure \ref{rdradscan} we can also detect the faint leading wake of Prometheus in R/2004 S2, which complicates any attempts to identify resonant structures that might exist there. The Prometheus ringlet shows a distinct offset in radius and a variation in brightness on either side of the central images corresponding to when Prometheus passed through the longitudinal region captured in the image-movie. Additionally, radial scans of images taken after Atlas passed by (right after Prometheus) show a slightly wider R/2004 S1 region.

As described above, the various trends in reflectance caused by different changes in lighting geometry within an image sequence and between sequences obtained over the course of the mission complicate efforts to quantify the amplitudes of these structures. To best facilitate comparisons between observations without compromising the integrity of the data we use a simple but robust normalization procedure. 
While a third-order polynomial fit to individual radial scans can reasonably remove these trends, it cannot without also adversely altering the appearance of the subtle structures we are interested in. Hence, we found that the best solution to create a set of normalized profiles within an image sequence is to subtract a linear trend in reflectance with time at each radius across the mosaic. This accounts for variation in brightness across images taken from drifts in lighting geometry during an image sequence while preserving any periodic variations due to the resonant structures. The ideal result is a vertically flat Roche Division map (no brightness slope with time) at radii where no structures occur. This method can cause issues near the edges of the A ring, F ring, and Prometheus ringlet, so we limit the normalization to the region of the Roche Division stretching from $137,000 - 139,100$ km. Finally, for each Roche Division observation we determine the amplitude of the $m=-3$ longitudinal brightness variations by fitting the data in a radius vs. phase (see Equation \ref{radii}) mosaic to a sinusoid at every radius. The brightness amplitude at each radius is normalized to a fractional amplitude by comparing it to the average brightness level of the Roche Division at that radius for each observation. We present corrected maps of the Roche Division from every data set (or combination of data sets) included in Table \ref{imaging data} whose duration is longer than $\sim7$ hours in Appendix \ref{map appendix}.

For an example, in Figure \ref{example fracamp} we show the normalized Roche Division map and fractional variation in ring brightness vs. radius during Rev 281. This observation sequence had particularly favorable viewing geometry (low $B$ and high $\alpha$) and so provides the clearest example of the resonant structure in the Roche Division. The strongest structure peaks around $137,420$ km, near the edge of the A ring with a pattern speed of $798.9$ $^{\circ}$day$^{-1}$. Similarly strong patterns are present near this location both in 2006 and since 2016.

In general, the structures in the Roche Division occur in a region from just outside the edge of the A ring to the R/2004 S2 ringlet. In certain observations, bright and dark diagonal banding almost appears to cover this entire range of nearly $2,000$ km as one structure. In fact, diagonal bands in the mosaic for Rev 281 of Figure \ref{example fracamp} can be traced from $137,200-138,100$ km and only show possible interruptions around $137,700$ km where Atlas orbits. Other observations show more distinct substructures at each radial region. We provide a brief summary of the lengthier observations in Figure \ref{summary} and more detailed figures of each of these observations in Appendix \ref{map appendix}.

\begin{figure}[h]
\centering
\includegraphics[width=.93\linewidth]{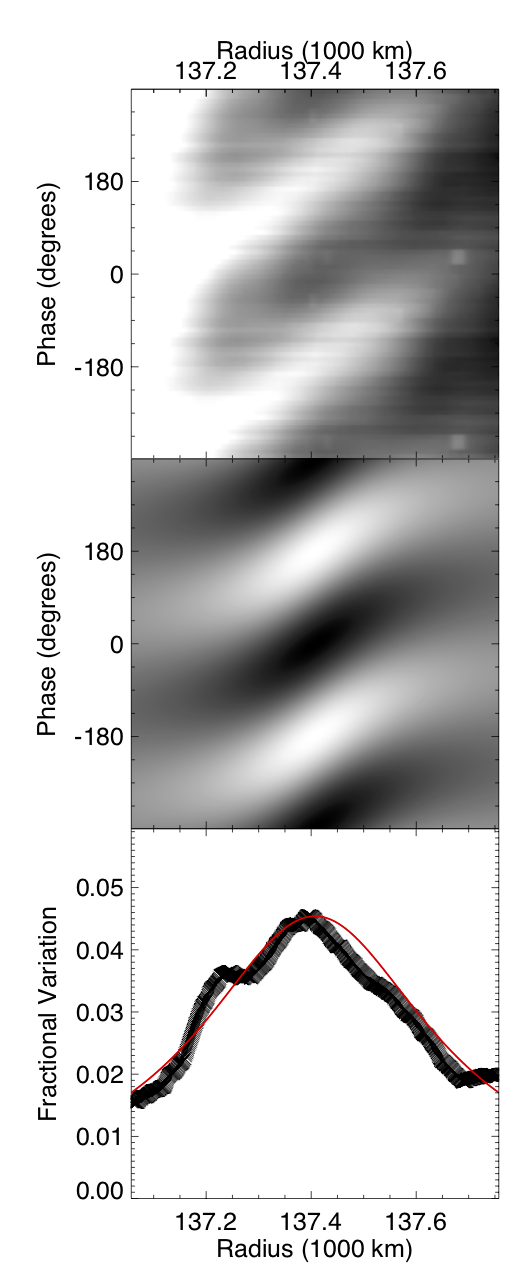}
\caption{The top panel is a small map of the high amplitude structure of Rev 281 from Figure \ref{example fracamp}, but with a pattern phase axis rather than time to conform with Equation \ref{particle density}. The middle panel is the model of the region made with Equation \ref{particle density}, after approximating $L$ as the full width at half maximum from a Lorentzian fit (red curve) to the bottom panel showing the fractional change in brightness across the structure at each radius. \label{model}}
\end{figure}

\begin{figure*}[t]
\centering
\includegraphics[width=.95\linewidth]{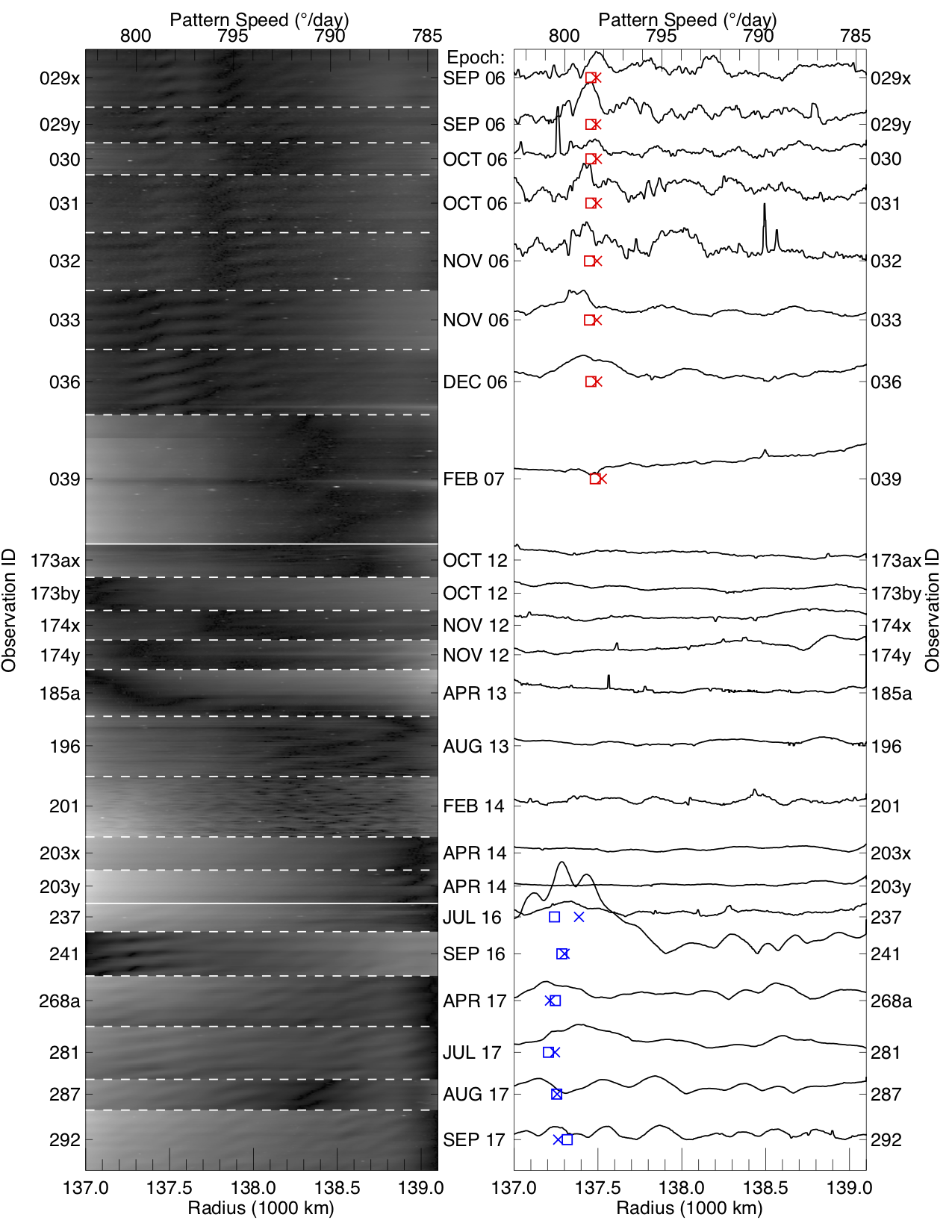}
\caption{This plot is a summary of Roche Division maps and their $m=-3$ fractional amplitudes of observations taken over the course of the \textit{Cassini} mission whose durations are 7 hours or longer. Each map is scaled to the same time scale (height), where for example Rev 196 is about one local orbital period ($\sim 14.5$ hours). Subsequent observations are stacked on top of one another sequentially in time, marked by the month and year the observation took place, and delineated by dashed lines or a solid line to separate out the main epoch ranges. The curves in the right portion of the figure are measures of the fractional amplitudes $\frac{amplitude\ of\ brightness\ variation}{average\ brightness}$ at each radius. These curves show the abundance of peaks occurring when appropriate northern (blue) or southern (red) SKR (X) and MAG(square) periods may be affecting the region. See Appendix \ref{map appendix} for larger plots of each observation. \label{summary}}
\end{figure*}

\section{Results}
We can characterize the Roche Division patterns using the radius of the peak amplitude and the equivalent pattern speed given by Equation \ref{exppat}, along with the structure's peak amplitude and width. Figure \ref{model} is a zoomed in look at the strongest structure of Rev 281 from Figure \ref{example fracamp}. The top panel is a mosaic of the observations and the middle panel is a mosaic modeled using Equation \ref{particle density}. After fitting the actual fractional brightness variations with a Lorentzian in the bottom plot, we can constrain the damping length $L$ (here, the half width at half maximum), the fractional amplitude $A$, and calculate the resonance strength $\beta$ at the exact resonance ($\delta r=0$) using Equation \ref{amplitude}. We can then use these parameters to estimate properties of the perturbation. 

\begin{deluxetable*}{cccccccccc}
\tablecaption{Peak pattern speeds and perturber strengths \label{patspeeds}}
\tablewidth{0pt}
\tablehead{
\colhead{Image} &
\colhead{Date} &
\colhead{Radius} &
\colhead{Pattern} &
\colhead{Pattern} &
\colhead{Peak fractional} &
\colhead{Damping} &
\colhead{Resonance} &
\colhead{Effective} &
\colhead{Magnetic field} \\
\colhead{movie} & 
\colhead{Month} &
\colhead{$r$} & 
\colhead{speed $\Omega_p$} & 
\colhead{period} & 
\colhead{amplitude} & 
\colhead{length $L$} & 
\colhead{strength $\beta$} & 
\colhead{perturber mass} &
\colhead{oscillation amp.} \\
\colhead{ID} & 
\colhead{Year} &
\colhead{(km)} & 
\colhead{($^{\circ}$day$^{-1}$)} & 
\colhead{(hours)} & 
\colhead{$A$} &
\colhead{(km)} &
\colhead{(km$^2$)} &
\colhead{$M_{Mimas}$} &
\colhead{$B_{\theta0}$ (nT)} \\ 
}
\startdata
029x & SEP 2006 & 137350 & 799.5 & 10.81 & 0.031 & 40 & 50 & 0.1 & 1.0 \\ 
029x & SEP 2006 & 137500 & 798.3 & 10.82 & 0.051 & 80 & 290 & 0.4 & 6.2 \\ 
029x & SEP 2006 & 138180 & 792.4 & 10.90 & 0.045 & 80 & 310 & 0.5 & 6.4 \\ 
029y & SEP 2006 & 137320 & 799.8 & 10.80 & 0.039 & 30 & 50 & 0.1 & 1.0 \\ 
029y & SEP 2006 & 137440 & 798.7 & 10.82 & 0.070 & 70 & 330 & 0.5 & 7.1 \\ 
030 & OCT 2006 & 137470 & 798.5 & 10.82 & 0.032 & 70 & 140 & 0.2 & 2.9 \\ 
031 & OCT 2006 & 137420 & 798.9 & 10.82 & 0.063 & 60 & 240 & 0.4 & 5.1 \\ 
033 & NOV 2006 & 137380 & 799.3 & 10.81 & 0.048 & 110 & 620 & 0.9 & 13.3 \\ 
036 & DEC 2006 & 137440 & 798.8 & 10.82 & 0.042 & 190 & 1510 & 2.3 & 32.0 \\ 
036 & DEC 2006 & 138040 & 793.6 & 10.89 & 0.022 & 120 & 310 & 0.5 & 6.5 \\ 
237 & JUL 2016 & 137320 & 799.8 & 10.80 & 0.024 & 160 & 630 & 0.9 & 13.4 \\ 
241 & SEP 2016 & 137120 & 801.6 & 10.78 & 0.099 & 130 & 1600 & 2.4 & 34.2 \\ 
241 & SEP 2016 & 137290 & 800.1 & 10.80 & 0.149 & 110 & 1660 & 2.5 & 35.4 \\ 
241 & SEP 2016 & 137430 & 798.8 & 10.82 & 0.129 & 130 & 2090 & 3.1 & 44.2 \\ 
268a & APR 2017 & 137230 & 800.6 & 10.79 & 0.029 & 170 & 800 & 1.2 & 17.1 \\ 
268a & APR 2017 & 138140 & 792.7 & 10.90 & 0.016 & 110 & 210 & 0.3 & 4.3 \\ 
268a & APR 2017 & 138580 & 788.9 & 10.95 & 0.028 & 120 & 420 & 0.6 & 8.6 \\ 
281 & JUL 2017 & 137420 & 798.9 & 10.81 & 0.044 & 270 & 3200 & 4.8 & 67.8 \\ 
281 & JUL 2017 & 137920 & 794.6 & 10.87 & 0.021 & 240 & 1150 & 1.7 & 23.9 \\ 
281 & JUL 2017 & 138600 & 788.7 & 10.95 & 0.027 & 120 & 390 & 0.6 & 7.9 \\ 
287 & AUG 2017 & 137140 & 801.4 & 10.78 & 0.026 & 100 & 270 & 0.2 & 5.7 \\ 
287 & AUG 2017 & 137540 & 797.9 & 10.83 & 0.024 & 120 & 330 & 0.5 & 6.9 \\ 
287 & AUG 2017 & 137840 & 795.3 & 10.86 & 0.030 & 110 & 380 & 0.6 & 8.0 \\ 
292 & SEP 2017 & 137250 & 800.4 & 10.79 & 0.022 & 80 & 130 & 0.2 & 2.7 \\ 
292 & SEP 2017 & 137550 & 797.8 & 10.83 & 0.023 & 60 & 80 & 0.1 & 1.7 \\ 
292 & SEP 2017 & 137870 & 795.0 & 10.87 & 0.024 & 76 & 140 & 0.2 & 2.9 \\ 
\enddata
\tablecomments{Using a Mimas mass of $3.7493 (\pm0.0031) \times10^{19}$ kg \citep{2006AJ....132.2520J}. The error of these measurements are dominated by systematic offsets that are difficult to quantify a priori.}
\end{deluxetable*}

\begin{figure}[t]
\includegraphics[width=\linewidth]{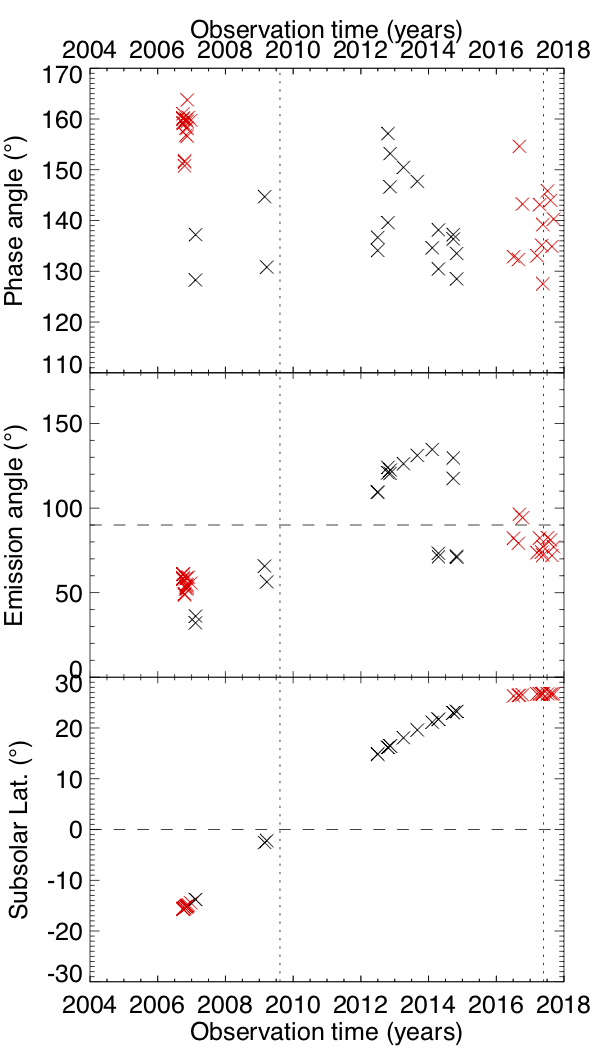}
\caption{These plots show the phase angle (top), emission angle (middle) of all Roche Division observations in Table \ref{imaging data}, and Saturn's subsolar latitude (bottom) versus the time of each observation. The red or dark symbols correspond to observations where the longitudinal brightness variations were visible in the Roche Division. The vertical dotted lines mark Saturn's equinox in 2009 and solstice in 2017. The horizontal dashed line marks $0^{\circ}$ and $90^{\circ}$ in the relevant plots. The lack of brightness variation detections in many observations cannot easily be attributed to their photometric angles. \label{lighting geometry}}
\end{figure}

In Table \ref{patspeeds} we present the spectrum of the strongest patterns in observation sequences longer than half a local orbital period ($\sim7.25$ hours) by noting their radii and the corresponding pattern speeds (and equivalent period). We also provide the relevant resonance strength parameters and estimate an effective perturber mass using Equation \ref{forced eccentricity} and magnetic field oscillation amplitude using Equation \ref{bamp}. The exact particle properties of the Roche Division are unknown at this time and are beyond the scope of this work. To provide an order of magnitude estimate of the magnetic oscillation amplitude $B_{\theta0}$, we can re-express the charge-to-mass ratio as $3\epsilon_0\Phi/\rho_gs^2$, where $\epsilon_0$ is the permittivity of free space, and assume a particle size $s=5$ $\mu$m \citep{1998DPS....30.1705S}, density $\rho_g=1$ g$/$m$^3$, and electrostatic potential $\Phi=10$ Volts \citep{2012Icar..219..498F}, consistent with prior estimates for dusty particles near the main rings \citep{2013Icar..225..446M}.

Both the peak amplitudes and widths of these structures vary dramatically among the various observations. 
For example, considering just the innermost structure we can see that the resonance strength appears to vary by over an order of magnitude. Observations like 033, 036, 268a, and 281 (see Figures \ref{rev033}, \ref{rev036}, \ref{rev268a}, and \ref{rev281}) have a single radially broad structure between Atlas and the A ring, while others, 029x, 241, 287, and 292 are made up of multiple narrower structures (see Figures \ref{rev029x}, \ref{rev241}, \ref{rev287}, and \ref{rev292}). This may be related to the strength of the rotating PPO current systems. Alternatively, some of these changes could be because the observations were obtained with different lighting geometries, which can affect the appearance of these structures. For instance, observations taken at different phase angles are sensitive to different particle sizes due to their light scattering properties. 

On a global scale, in Figure \ref{summary} patterns can be seen in all the observations from 2006 and 2016-2017, and no similar patterns are detected in 2012-2014. Close examination of the lighting and viewing conditions reveal that these alone cannot explain the overall variable presence of the patterns. Figure \ref{lighting geometry} shows the photometric angles for each observation throughout the mission. The red X's correspond to the observations where we are able to detect periodic structures in the Roche Division. The majority of observations in 2006 were taken from very high phase angles ($\alpha>150^{\circ}$), but the remaining observations were taken at a large range of phase angles ($\alpha\sim130-160^{\circ}$). Lower phase angle observations ($\alpha<140 ^{\circ}$) are generally less likely to reveal the structures, and reveal them with less detail. However, we cannot attribute the lack of structure identification from 2009-2015 to a lack of appropriate phase angle viewing, because several Revs (173by, 174x, 174y, 185a, and 196) have phase angles $\alpha>140^{\circ}$ and cover a sufficient length of time (at least half of a local orbital period), conditions under which patterns could be seen in 2016-2017.

The best candidate for missing the structure due to lighting geometry might be due to a combination of phase angle and the emission angle. These structures are most obvious in sequences with particularly high phase angles ($>150^{\circ}$) or with emission angles closest to $90^{\circ}$ (i.e. low ring opening angles $|B|<10^{\circ}$). While there isn't an obvious missed opportunity of comparable lighting geometry during the 2009-2015 epoch of no Roche Division structures, we should examine this possibility closely. Notice in the central plot of Figure \ref{lighting geometry}, we cannot compare sequences having the same emission angle between each of the three epochs, as we are able to with phase angles. Emission angles $e<90^{\circ}$ correspond to observations taken from the north side of Saturn's ring-plane. During the first observations of the Roche Division, in late 2006, Saturn's southern hemisphere was illuminated by the Sun, so these observations with emission angles $e \sim50-60^{\circ}$ were taken from the unlit side of the rings. Similarly, several observations after equinox were taken from the unlit side of the rings, but from the south side of the ring plane. However, this spread in emission angles ($e \sim 110-140^{\circ}$) does cover a similar range in absolute ring-plane opening angles $|B|=|90^{\circ}-e|\approx35^{\circ}$ on the unlit side of the rings. If the structures were present all the time, and only visible under optimal lighting geometry, there is no reason that we shouldn't see the structures in Rev 173by and 174x when we can see them in Rev 031 having nearly identical lighting geometry (see lighting geometry in Table \ref{imaging data} and maps in Appendix \ref{map appendix}). Therefore, we conclude the presence of the Roche Division structures must be due to changes in the frequencies of the forces perturbing them.

\section{Discussion}

\subsection{Pattern speeds and locations}

\begin{figure}[t]
\includegraphics[width=\linewidth]{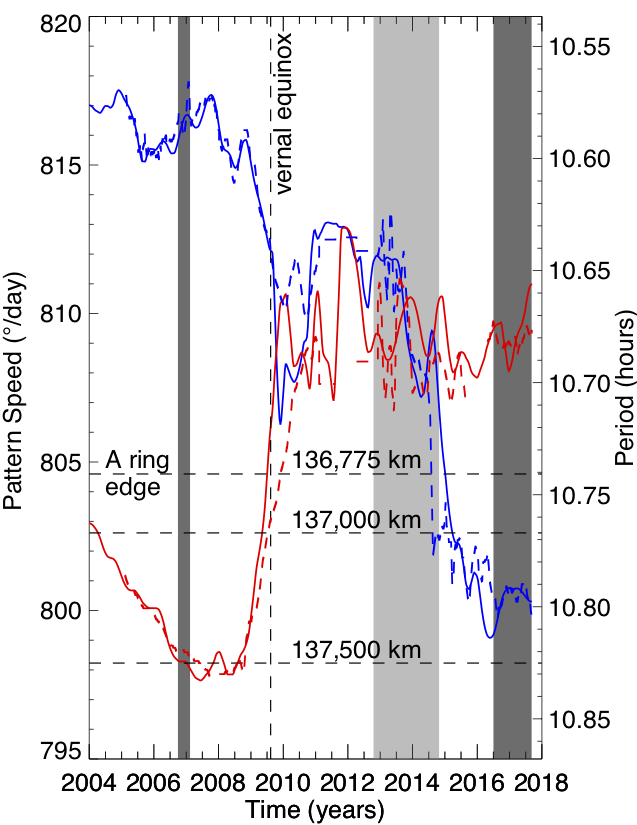}
\caption{This plot shows the rotational modulation rates of the Southern (red) and Northern (blue) hemispherical components of the SKR \citep{2018GeoRL..45.7297Y} (solid) and MAG \citep{2018JGRA..123.3859P} (dashed) versus time. Light horizontal dashed lines mark the effective edge of the A ring edge (near $805$ $^{\circ}$day$^{-1}$) , and the pattern speeds for reference radii at $137,000$ and $137,500$ km. A vertical dashed line marks the vernal equinox. The vertical gray bands encompass periods spanned by the observations in Table \ref{imaging data}. The darker bands around 2006 and 2016-2017 reflect the observation epochs when we have clear detections of brightness variations in the Roche Division. The lighter band covers the observation epoch around 2012-2014 when there wasn't a single detection of brightness variations in the Roche Division. This suggests that the presence of periodic brightness variations in the Roche Division is correlated with times when there are suitably long magnetospheric periods. \label{sls5rates}}
\end{figure}

The seasonal appearance of resonant structures in Saturn's Roche Division fits in well with the observed seasonal variations in Saturn's PPO rotational modulations. This is demonstrated in Figure \ref{sls5rates}, which shows the rotation speeds of the northern (blue) and southern (red) PPO systems derived from the SKR (solid line) and MAG (dashed line) data over the \textit{Cassini} mission \citep{2016JGRA..12111714Y,2018GeoRL..45.7297Y,2016JGRA..121.9829P,2018JGRA..123.3859P}. An equivalent scale of PPO periods is shown on the right hand vertical axis. As briefly indicated in Section \ref{intro}, the rates were well separated early in the mission under Saturn southern summer conditions at $816^{\circ}$day$^{-1}$ for the weaker northern system and $\sim800^{\circ}$day$^{-1}$ for the stronger southern system, before converging towards near-equal rates with near-equal amplitudes across vernal equinox in mid-2009. An interval of variable but near-equal rates with variable amplitudes then ensued between mid-2010 and mid-2013 \citep{2013JGRA..118.3243P,2015Icar..254...72F}, though the northern rate $\sim812^{\circ}$day$^{-1}$ generally remained larger than the southern rate $\sim809^{\circ}$day$^{-1}$, before it coalesced at $\sim808^{\circ}$day$^{-1}$ between mid-2013 and mid-2014. The rates then separated again, with the now-stronger northern system slowing to $\sim800^{\circ}$day$^{-1}$ while the weaker southern system remained near $\sim809^{\circ}$day$^{-1}$. The Roche Division structures reported by \citet{2009Icar..202..260H} were observed during the initial interval of well-separated rotation rates, as indicated by the left hand dark gray band in Figure \ref{sls5rates}, when the period of the stronger southern system was $\sim798^{\circ}$day$^{-1}$. However, during the extended post-equinox interval of near-equal northern and southern system rates $\sim810^{\circ}$day$^{-1}$, all observations of the Roche Division show no evidence of such periodic brightness variations, as indicated by the central gray band in Figure \ref{sls5rates}. (See Appendix \ref{map appendix} for the individual maps and fractional amplitudes of these observations.)


During the time between early 2009 and late 2014 when we observe no ring structures, both PPO rates are above $805^{\circ}$day$^{-1}$ \citep{2015Icar..254...72F,2016JGRA..12111714Y,2016JGRA..121.9829P}. Thus, an appropriately commensurate 3:4 OLR with a pattern speed greater than $805$ $^{\circ}$day$^{-1}$ would be located interior to the outer edge of Saturn's A ring and unsuitable to produce resonant structures in the Roche Division. Horizontal dashed lines in Figure \ref{sls5rates} show the pattern speed at $\sim805$ $^{\circ}$day$^{-1}$ for such a resonance at the outer edge of the A ring, as well as at reference radii of $137,000$ and $137,500$ km in the region outside the edge of the A ring. We first observe the reappearance of the Roche Division structures in Rev 237 on July 1st, 2016 and they are apparent in all later observation sequences. The closest prior observation to Rev 237 is Rev 209 in October 2014 just after the rate of the northern component began to slow down after the interval of coalesced periods.


Additionally, during 2006 and 2016-2017 the PPOs of the stronger systems have pattern speeds that coincide with the location of the largest fractional variations in the Roche Division. This is most notable between the observations in 2006 when the southern PPO rate slowed to below $799$ $^{\circ}$day$^{-1}$ and the observations in 2017 when the northern PPO rate was around $800$ $^{\circ}$day$^{-1}$. In the 2006 observations the location with the highest fractional brightness variations is further out into the Roche Division around $137,400-137,500$ km while in 2017 some strong structures appear even closer to the A ring around $137,100-137,300$ km (See Figure \ref{summary}). This suggests that the structures in the inner Roche Division are tied to the presence of the variable PPO rates. 

\begin{figure*}[ht]
\includegraphics[width=\linewidth]{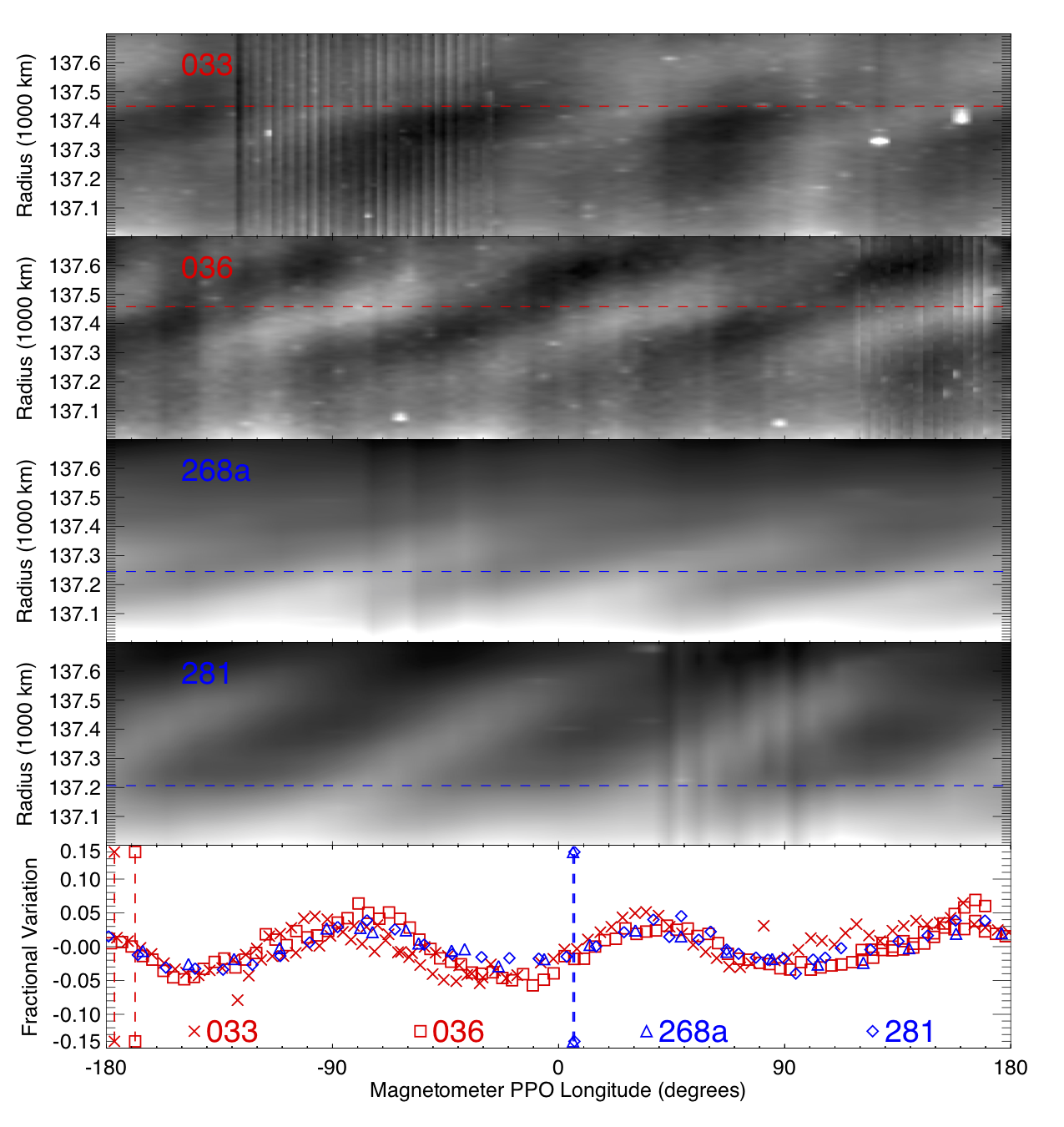}
\caption{The top four panels contain maps of four Roche Division observations using the MAG PPO longitude system, see Appendix \ref{magsystem} \citep{2018JGRA..123.3859P}. Dashed lines mark the resonant radius of the southern (red) or northern (blue) PPO period at the time of the ring observations. In the lower plot are the brightness profiles of the ring map at the resonant radius marked above. The vertical axis in this plot represent the fractional variation above or below the mean brightness at the resonant radius. The vertical dashed lines in the bottom plot mark the longitude of the maximum SKR emission in the MAG PPO system for each observation \citep{2018GeoRL..45.7297Y}. The innermost Roche Division structure appears to be in sync with the PPO longitude. \label{magphase}}
\end{figure*}

In fact, we find that the innermost Roche Division structure is in sync with the PPOs. In Figure \ref{magphase} we show four of the best observations of the mission mapped to the appropriate MAG PPO longitude system. In their determination of the PPO periods the magnetometer team also defines a longitude system and tracks the phases of both PPO systems \citep{2018JGRA..123.3859P}. For each ring observation shown we determine the longitude of each ring profile in the appropriate north or south PPO system at the time the image was taken. See Appendix \ref{magsystem} for a complete description of the PPO longitude systems and how we bring our ring observations into this system to create Figure \ref{magphase}. The effective MAG PPO resonant radius for the northern (blue) and southern (red) systems are marked with dashed lines on the maps. The fractional variation of the brightness profile at each of these radii is plotted below with the same color scheme for north/south and a different symbol for each ring observation. The SKR maximum emission in this MAG longitude system is marked by the vertical dashed lines with appropriate symbols on each end \citep{2018GeoRL..45.7297Y}. The minimum fractional variations of the ring pattern appears to be consistently at quadrature with the PPO system that is responsible for the resonance in the Roche Division at any given time. The zero point of the PPO longitude system is defined where the oscillating magnitude of $B_\theta$ is a minimum in the northern PPO system and where $B_\theta$ is a maximum in the southern PPO system. Strangely, we do not see a $180^{\circ}$ offset between the northern and southern PPO affected ring structures. Nonetheless, the coherence of these structures at the drifting resonant radius is suggestive of a clear relation between the PPO and the inner Roche Division.


The connection between the outer Roche Division structures, with slower expected pattern speeds, and the PPOs is less clear. We do not understand how this phenomena can perturb an almost $2000$-km-wide radial span in the Roche Division. It is possible that the outer structures are due to other longer periods, but these are not seen in the MAG and SKR data. Also, if they are due to different periodic forces they would not necessarily disappear at the same time as the faster PPO signals. Our detection of resonant structures so far from the exact PPO resonance could be due to unexpectedly long propagation from the resonant radius. 

Perhaps these extensive dusty structures in the Roche Division are generated by similar phenomena as the spokes. The dusty spokes in Saturn's B ring, which extend over a radial range up to $10,000$ km \citep{1981Sci...212..163S}, are also related to the PPO periods \citep{2013Icar..225..446M}. The spokes in the B ring and the resonant Roche Division structures certainly have some connection, but the dust making up the B ring spokes must be levitated above and below the numerous larger ring particles \citep{2006Sci...311.1587M}, introducing added complexity to the dusts' response to resonant forcing from the PPO near co-rotation. Indeed one might wonder why spoke-like structures do not form in the outer A ring when the PPO periods are both $\sim810^{\circ}$deg$^{-1}$. There may be a different response to the PPO by dust within dense rings and dust in dust-dominated rings preventing propagation across this boundary. At times when the PPO resonant radius falls directly in a dust-dominated ring, the structures may be able to propagate across the dusty region, with some interruptions from embedded moons and other gravitational influences. This is an idea that will need to be explored more explicitly in a future work.

\subsection{Perturbation strengths \label{strength disc}}
Additional insights into the sources of these perturbations can be extracted from their amplitudes. The structures' typical peak amplitudes are $2-7\%$ of the local mean Roche Division brightness with damping lengths of $\sim50-300$ km (see Table \ref{patspeeds}). If these structures were generated by gravitational forces, then hypothetical perturber masses cover a wide range, from nearly five Mimas masses to just a tenth of a Mimas mass. Again, the innermost structure is perturbed most heavily. However, we should recall that these mass estimates are the result of an incomplete phenomenological model. Applying this simple dissipation model to the 8:7 Mimas ILR in the G ring, we measure the mass of Mimas to be $2.5$ times the actual Mimas mass, so these results are likely just an order of magnitude estimate of an actual perturber mass \citep{2009Icar..202..260H}. On the other hand, if these perturbations are due to oscillations in the magnetic field, then our estimates of the magnetic field oscillation amplitude $B_{\theta0}$ are within an order of magnitude of the \citet{2013JGRA..118.3243P,2016JGRA..121.9829P,2018JGRA..123.3859P} results. This provides further evidence that the perturbing forces are electromagnetic rather than gravitational. 

Considerations of the expected gravitational effects of other moons on the Roche Division confirms that the perturbations responsible for producing these patterns are very strong and likely non-gravitational. Saturn's moons Janus and Epimetheus are similar-mass co-orbital satellites that exhibit horseshoe orbits with an 8 year period \citep{1981Icar...47...97H}. The Janus/Epimetheus 7:6 ILR is located on the edge of the A ring and shapes its 7-lobed wavy edge \citep{2016Icar..279..125E}. This resonance, along with the combined torques from several satellite resonances in the outer region of the A ring, confines Saturn's main rings \citep{2017ApJS..232...28T}. While the exact location of the Janus and Epimetheus resonances drift slightly over their 8 year orbital swap, the 8:7 ILR always occurs in the same part of the outer Roche Division and so might be expected to shape that region. However, there are no obvious 8-lobed or canted structures moving at the moons' orbital rates in any of the image-movies presented here.

Instead, the brightness variations of a 3:4 OLR appear to dominate this location (see Rev 281), but the structure is consistently not as clean as those in other radial locations. The Janus/Epimetheus 8:7 ILR should be even stronger than its 7:6 ILR, but the lack of an 8 lobed structure similar to that found in the G ring at the Mimas 8:7 ILR is consistent with high gravitational masses needed to produce the 3:4 resonant structures. Thus, the strength of the 3:4 OLR with Saturn's PPOs must be even greater if it is the dominant perturbing force in this region. This further reinforces the conclusion that these structures are the result of electromagnetic perturbations associated with Saturn's rotating PPO systems. In fact, our measured values of $\beta$ for the 3:4 OLR structures closest to the Janus/Epimetheus 8:7 ILRs are $420$ km$^2$ and $390$ km$^2$ for Revs 268a and 281. These are more than 10 times larger than those predicted for the Janus 8:7 ILR using the mass found by \citet{2008AJ....135..261J}.

As mentioned earlier, leading Prometheus wakes are also observed in the R/2004 S2 ringlet near this location adding to the competing perturbing forces. One might also expect, due to the close proximity of the Roche Division dust bands to Atlas and Prometheus, that gravitational satellite wakes should dominate the Roche Division's appearance. Upon close inspection of Figure \ref{rdradscan}, we can see the wakes leading Prometheus in the outer band R/2004 S2. Satellite wakes have azimuthal wavelength $\lambda=3\pi\delta a$ related to the radial separation of the satellite and the ring material $\delta a=|a-a_s|$ \citep{1985ApJ...292..276C,1986Icar...66..297S,2016AJ....152..211C}. This wavelength is $\sim2^{\circ}$ in longitude compared to the $120^{\circ}$ variations of the 3:4 OLR. We have not detected any evidence of satellite wakes in the dust around Atlas. Some of the material in Atlas' vicinity could be co-rotating like the Prometheus ringlet \citep{2017Icar..281..322H} or the Encke Gap ringlets \citep{2013Icar..223..252H}, but we have not seen any evidence supporting either arrangement. 



\section{Summary and conclusions}
In this work we have tracked the presence of dusty periodic structures in the Roche Division throughout the course of the \textit{Cassini} mission. We find these structures are present when the 3:4 OLR with Saturn's seasonally-varying magnetospheric periodicities are in this region. By analyzing the resonant response of the rings and applying a simple model of an oscillating magnetic field we are able to estimate oscillation amplitudes comparable to those observed with \textit{Cassini's} magnetometer instrument. 

It is evident that electromagnetic forces dominate the perturbations in the Roche Division. The effect of nearby small satellites and satellite resonances are mostly washed out by the dominant influence of the 3:4 OLR with the rotating magnetic field perturbation. The varying nature of the magnetospheric periodicities is coincident with the presence of Roche Division structures, providing further evidence of this association. Further investigation of the magnetospheric observations connected to the Roche Division and investigations of other dusty ring structures could resolve the issue of how these patterns span a wide range of radii. This could be relevant to studies of other expansive dust systems, like protoplanetary disks.

As previously discussed, the ring regions where the most analogous structures have been detected are the G ring and D ring. While the dusty resonant structure in the G ring is known to be caused by the 8:7 ILR with Mimas, the D ring seems to be perturbed by the same phenomena as the Roche Division, but at the 2:1 ILR with the PPO periods \citep{2009Icar..202..260H}. However, the D ring spans a much wider radial range of pattern speeds matched by various phenomena associated with Saturn's rotation. While the Roche Division only covers the slowest rotation periods of the PPO, the D ring covers the full range of these seasonal magnetospheric periods and the full range of Saturn's latitudinal wind speeds. In fact, the prevalence of resonant structures resulting from 2:1 ILRs warrants a separate work. However, we note that the faint region of the D ring between the D72 and D73 ringlets appears to be responding to the same phenomena we report on here for the Roche Division. 

The presence of ring structures at these particular resonances with the PPO makes sense because the D ring and Roche Division contain large populations of dust. The remaining resonances with various $|m|$ live mostly in the dense A, B, and C rings. The only other such PPO resonance falling in a dusty ring is the 2:3 OLR ($m=-2$) located near the inner edge of the Janus-Epimetheus ring ($\sim149,000$ km from Saturn's center). This location has not been monitored at an extent comparable to D ring or Roche Division, but our brief investigation of the limited data revealed no obvious structures. As mentioned previously, the B ring spokes may be an expression of the PPO perturbations surrounding PPO co-rotation. Future studies should examine the connections between the B ring spokes and these patterns in the D ring and Roche Division, and why there are no dusty spokes or similar structures around other PPO resonances like the 3:2, 4:3, and 5:4 ILR, or 4:5 OLR.

\acknowledgements
This work was supported by NASA through the Cassini Data Analysis Program NNX15AQ67G. Work at the University of Leicester was supported by STFC Consolidated Grant ST/N000749/1. We thank Georg Fischer for helpful comments on an earlier version of this manuscript. We are grateful for helpful conversations with Douglas P. Hamilton concerning Lorentz resonances and quantifying their strengths. We also thank the Cassini project and imaging team for acquiring the data for this work. We would also like to thank two anonymous reviewers for their helpful suggestions that ultimately improved the clarity and consistency of this manuscript.


\appendix
\section{Faint ring response model \label{derivations}}
Here, we derive the model developed by \citet{2009Icar..202..260H} with some slight corrections. We can approximate the dynamics of ring particles near a Lindblad resonance with a driven harmonic oscillator. Let $x=r-a$ represent a ring particle's radial excursions from semi-major axis $a$. The ring particle has a mean motion $n$ close to that of the exact resonance $n_r$ and its equation of motion is given by
\begin{equation}
\ddot{x}=-n^2x-B\cos(n_rt)
\end{equation}
where $B$ is the strength of the driven oscillations. The solution to the equation of motion is
\begin{equation}
x=\frac{B}{n_r^2-n^2}\cos(n_rt).
\end{equation}
We can further approximate
\begin{equation}
n_r^2-n^2\approx\frac{3GM}{a_r^4}\delta a
\end{equation}
where $n^2=GM_S/a^3$, $a_r$ is the resonant radius, and $\delta a=a-a_r$.
Now, we can rewrite $x$ as
\begin{equation}
x=r-a=\frac{\beta}{\delta a}\cos{\phi'}
\end{equation}
where $\beta=B(3GM_S/a_r^4)^{-1}$ is the resonance strength and $\phi'=n_rt$ is related to the phase $\phi$ used in Equation \ref{radii} after approximating $\frac{m}{m-1}\approx\frac{n}{n_r}$. The distinction between $\phi$ and $\phi'$ is not noted by \citet{2009Icar..202..260H}, but is important to address. The phase parameter used previously, $\phi=m(\lambda-\Omega_p(t-t_0)-\delta_m)$ or more simply $\phi=m(\lambda-\lambda_s)$, varies differently than $\phi'$ with time if one considers the nature of our ring observations. In the ring image-movies we stare at one longitudinal region as time increases ($\dot{\lambda}=0$), so $\phi$ is actually decreasing as time passes in our image sequence observations. Alternatively, one can clearly see that $\phi'$ should increase with time if $n_r>0$. This will become important to consider when applying this simple model to the actual data and the direction the structure should be tilted, particularly for Figure \ref{model}.

After adding a damping term the equation of motion becomes:
\begin{equation}
\ddot{x}=-n^2x-\frac{1}{\tau}\dot{x}-B\cos(n_rt) \label{eom2}
\end{equation} 
where $\tau$ is the damping time of the radial excursions. The solution to the new equation of motion is
\begin{equation}
x=x_0\cos(n_rt-\phi_0')
\end{equation}
where
\begin{equation}
x_0=\frac{B}{(n_r^2-n^2)\cos{\phi_0'}+(n_r/\tau)\sin{\phi_0'}} \label{x0}
\end{equation}
and
\begin{equation}
\tan{\phi_0'}=\frac{n_r/\tau}{n_r^2-n^2} \label{tan}.
\end{equation}
The radii of particle streamlines can then be expressed as
\begin{equation}
r=a+x=a+x_0\cos(\phi'-\phi_0')=a+x_0(\cos{\phi'}\cos{\phi_0'}-\sin{\phi'}\sin{\phi_0'}).
\end{equation}
After finding the expressions for $\cos{\phi_0'}$ and $\sin{\phi_0'}$ from Equations \ref{x0} and \ref{tan} we can define the radius as
\begin{equation}
r=a+\frac{\beta}{(\delta a)^2+L^2}(\delta a\cos{\phi'}-L\sin{\phi'}) \label{radmodel}
\end{equation}
where $L=a_r/(3n_r\tau)$ is the damping length. This streamline model results in a gradual shift of pericenter locations across the resonance, which is responsible for the canted brightness variations peaking at the exact resonance due to increased density of particles.

We assume that all particles are evenly distributed in semi-major axis and phase, and the phase space number density is constant $\rho(a,\phi)=\bar{\rho}$. Or, as a function of radius and phase
\begin{equation}
\rho(r,\phi)=\frac{\bar{\rho}}{dr/da}
\end{equation}
where we assume the streamlines do not cross (i.e. $dr/da$ never equals zero). In fact, the observed fractional variations in brightness are only a few percent at most, which is inconsistent with streamline crossing. Thus, the fractional variations in density are given by
\begin{equation}
\frac{\delta\rho(r,\phi)}{\rho}=\frac{\rho(r,\phi)-\bar{\rho}}{\bar{\rho}}\approx1-\frac{dr}{da}.
\end{equation}
We can then differentiate Equation \ref{radmodel} to get
\begin{equation}
\frac{dr}{da}=1-\frac{\beta}{[(\delta a)^2+L^2]^2}\left[\left((\delta a)^2-L^2\right)\cos{\phi'}-2\delta aL\sin{\phi'}\right]
\end{equation}
and, since $\delta r \approx \delta a$
\begin{equation}
\frac{\delta\rho(r,\phi')}{\rho}=\frac{\beta}{[(\delta r)^2+L^2]^2}\left[\left((\delta r)^2-L^2\right)\cos{\phi'}-2\delta rL\sin{\phi'}\right] \label{density model}
\end{equation}
where our amplitude of density variations is given by
\begin{equation}
A(r)=\frac{\beta}{(\delta r)^2+L^2}.
\end{equation}
The difference between Equation \ref{density model} and Equation \ref{particle density} is due to the subtle difference of $\phi'$ and $\phi$.

\section{Lorentz resonance strength \label{lorentz}}
The perturbation equations needed to solve for the Lorentz resonance strength were derived by \citet{1994Icar..109..221H}. We use the appropriate second-order expansion of the perturbation equations in their Table II. Specifically row three of the $g_{4,3}$ portion of Table II, because the resonant argument for the 3:4 OLR or effectively a 3:4 Lorentz resonance is $\varphi=-4\lambda+3\lambda_s+\dot{\varpi}$. The approximate expansions are a function of the ratio of magnetic field coefficients $\frac{g_{4,3}}{g_{1,0}}$ appropriate for the 3:4 resonance:
\begin{equation}
\left\langle \frac{de}{dt} \right\rangle = 2\left(1-\frac{n}{\Omega_S}\right)nL\frac{\sqrt{70}}{16}\frac{g_{4,3}}{g_{1,0}}\left(\frac{R_S}{a}\right)^3\sin{\varphi}
\end{equation}
and
\begin{equation}
\left\langle \frac{d\varpi}{dt} \right\rangle = -2\left(1-\frac{n}{\Omega_S}\right)\frac{nL}{e}\frac{\sqrt{70}}{16}\frac{g_{4,3}}{g_{1,0}}\left(\frac{R_S}{a}\right)^3\cos{\varphi}. \label{dapsedt}
\end{equation} 
Here, $\Omega_S$ is the rotational frequency of Saturn's magnetic field, $R_S$ is Saturn's equatorial radius, and $L$ is the ratio of the Lorentz force and the planetary gravitational force: 
\begin{equation}
L=\frac{q_g}{m_g}\frac{g_{1,0}R_S^3\Omega_S}{GM_S}. \label{forceratio}
\end{equation}
This force ratio is also dependent on Saturn's physical parameters, including mass $M_S$, the charge to mass ratio of the dust grains being perturbed $\frac{q_g}{m_g}$, and Saturn's dipole coefficient $g_{1,0}$. The stable solution requires $\left\langle\frac{de}{dt}\right\rangle=0$, therefore, $\sin(\varphi)=0$ and $\cos(\varphi)=\pm1$.

To solve for the resonance strength we follow the logic of \citet{hedman_2018}. Taking the time derivative of Equation \ref{resarg} we find:
\begin{equation}
\dot{\varphi}=m(n-n_s)-\left(n-\frac{d\varpi}{dt}\right)
\end{equation}
and effectively
\begin{equation}
\dot{\varphi_0}=m(n-n_s)-(n-\dot{\varpi_0})
\end{equation}
where $\dot{\varpi_0}$ is the secular apsidal precession rate due to Saturn's finite oblateness. Consider the difference of $\dot{\varphi}$ and $\dot{\varphi_0}$:
\begin{equation}
\dot{\varphi}-\dot{\varphi_0}=m(n-n_s)-\left(n-\frac{d\varpi}{dt}\right)-m(n-n_s)+(n-\dot{\varpi_0})=\frac{d\varpi}{dt}-\dot{\varpi_0}. \label{phidiff}
\end{equation} 
At the semi-major axis of a first-order resonance $a_r$ for a particular integer $m$, $\varphi$ is constant, and $\dot{\varphi}=0$. Additionally, for ring particles orbiting near the resonance, or small $\delta a=a-a_r$, we can approximate $\dot{\varphi_0}$ by expanding in terms of $\frac{\delta a}{a_r}$:
\begin{equation}
\dot{\varphi_0}=-n\frac{\delta a}{a_r}\left[\frac{3}{2}(m-1)+\frac{21}{4}J_2\left(\frac{R_S}{a_r}\right)^2\right]\approx -n\frac{\delta a}{a_r}\frac{3}{2}(m-1).
\end{equation}
Equation \ref{dapsedt} should also contain a $\dot{\varpi_0}$ term to account for Saturn's finite oblateness, so we can rewrite Equation \ref{phidiff} as:
\begin{equation}
0+n\frac{\delta a}{a_r}\frac{3}{2}(m-1)=-2\left(1-\frac{n}{\Omega_S}\right)\frac{nL}{e}\frac{\sqrt{70}}{16}\frac{g_{4,3}}{g_{1,0}}\left(\frac{R_S}{a_r}\right)^3\cos{\varphi}+\dot{\varpi_0}-\dot{\varpi_0}.
\end{equation}
We can now solve for a forced eccentricity $a_re_f=\frac{\beta}{\delta a}$ similar to Equation \ref{forced eccentricity}:
\begin{equation}
a_re_f=\frac{\sqrt{70}}{12}\left(\frac{n/\Omega_S-1}{m-1}\right)L\frac{g_{4,3}}{g_{1,0}}\left(\frac{R_S}{a_r}\right)^3\frac{a_r^2}{\delta a}, \label{gforcede}
\end{equation}
and so in this case
\begin{equation}
\beta=\frac{\sqrt{70}}{12}\left(\frac{n/\Omega_S-1}{m-1}\right)L\frac{g_{4,3}}{g_{1,0}}\left(\frac{R_S}{a_r}\right)^3a_r^2
\end{equation}
is a measure of the resonance strength due to an electromagnetic perturbation. After substituting the right hand side of Equation \ref{forceratio} for $L$:
\begin{equation}
\beta=\frac{\sqrt{70}}{12}\left(\frac{n/\Omega_S-1}{m-1}\right)\frac{q_g}{m_g}\frac{R_S^3\Omega_S a_r^2}{GM_S}g_{4,3}\left(\frac{R_S}{a_r}\right)^3.
\end{equation}
A complete estimation of the resonance strength should also include contributions from, $h_{4,3}$ and other $g_{j,3}$ and $h_{j,3}$ coefficients. Assuming the charge to mass ratio for the ring particles in the Roche Division used above we find an upper limit of $g_{4,3}$ on the order of a few hundred to a few thousand nanotesla. This is far greater than those measured for the higher order axis-symmetric coefficients \citep{2018Sci...362.5434D}. However, measurements of the small magnetic field oscillations associated with the PPO perturbation field resemble our results. To compare with these measurements, we develop a new approximation for the electromagnetic perturbations on dusty ring regions very similar to the formulation shown here (See Section \ref{magmodel}).

\section{PPO Longitude System \label{magsystem}}
\begin{figure*}[ht]
\centering
\includegraphics[width=.76\linewidth]{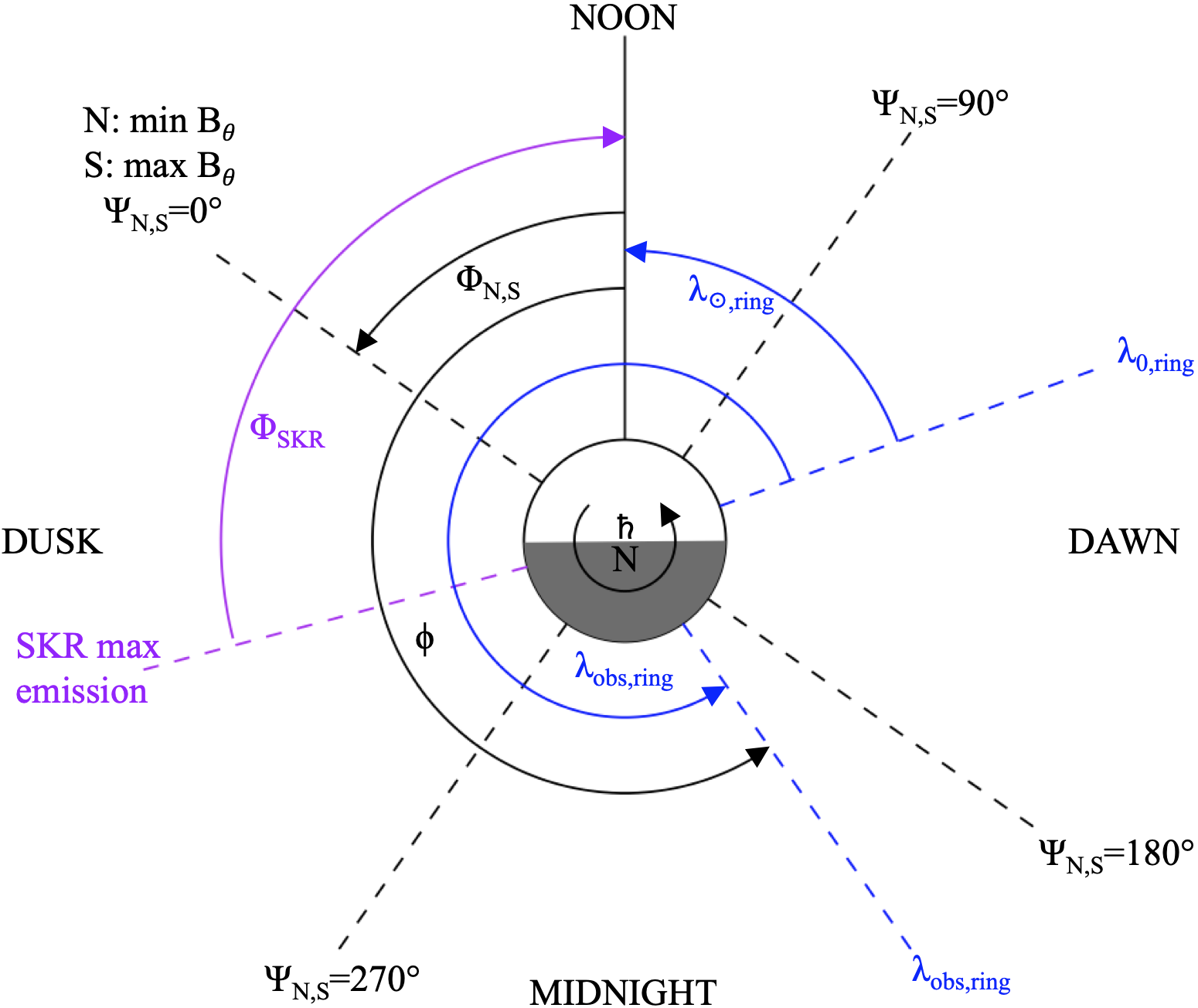}
\caption{A model of Saturn's PPO longitude system viewed from above the north pole, meant to enable a further understanding of Figure \ref{magphase}. The MAG PPO longitude system $\Psi_{N/S}$ is labeled in black, the ring inertial longitude system $\lambda_{ring}$ relative to the ascending node of the rings on J2000 is labeled in blue, and the SKR phase $\Phi_{SKR}$ is labeled in purple. The longitude of the Sun at Saturn noon is handily used to relate the angles between the various longitude systems. In the MAG PPO longitude system $\Psi_{N/S}=0^{\circ}$ corresponds to a minimum $B_{\theta}$ component in the northern hemisphere, and a maximum $B_{\theta}$ in the southern hemisphere. \label{pposystem}}
\end{figure*}

We use the \textit{Cassini} magnetometer PPO longitude system defined in \citet{2018JGRA..123.3859P} and detailed in Figure \ref{pposystem} to highlight our incorporation of ring observation longitudes. The PPO longitude of the northern or southern hemisphere $\Psi_{N/S}=\Phi_{N/S}-\phi$, where $\Phi_{N/S}$ are the PPO phases measured throughout the mission by the magnetometer team and $\phi$ is a longitude measured in the prograde direction from the sub-solar longitude. The PPO phases $\Phi_{N/S}$ are also measured from the sub-solar longitude in the prograde direction and define the zero point of the PPO longitude system $\Psi_{N/S}$. To get our ring observations into this longitude system we simply need to know the phase of the appropriate hemisphere's PPO $\Phi_{N/S}$ at the time of the observation and the angle of the observation $\phi$ relative to the sub-solar longitude. A ring observation's inertial longitudes $\lambda_{obs,ring}$ are measured in the prograde direction from a zero point $\lambda_{0,ring}$ relative to the ascending node of the rings on J2000. This means that all we need is the longitude of the sub-solar point in the ring longitude system $\lambda_{\sun,ring}$ to then calculate $\phi=\lambda_{obs,ring}-\lambda_{\sun,ring}$ needed to determine $\Psi_{N/S}$. Each ring image-movie observation captures one longitudinal region of the rings for several hours, but the ring system longitude is essentially identical. As the hours pass the images do however capture a spread of longitudes in the PPO system. This is how we are able to create Figure \ref{magphase}.

The SKR is similar to convert into the magnetometer PPO system. The SKR phases $\Phi_{SKR}$ track the location of maximum SKR emission relative to the sub-solar longitude \citep{2018GeoRL..45.7297Y}. In this case we can readily substitute $\phi=\Phi_{SKR}$ and determine the SKR North or South longitude in the appropriate MAG PPO system $\Psi_{N/S}$. Because the magnetic field and SKR oscillations have essentially the same period, the SKR maximum emission is fixed at a constant offset relative to the zero point of $\Psi_{N/S}$ at any given time.

\section{Roche Division Maps \label{map appendix}}
\begin{figure*}[ht]
\centering
\includegraphics[width=.76\linewidth]{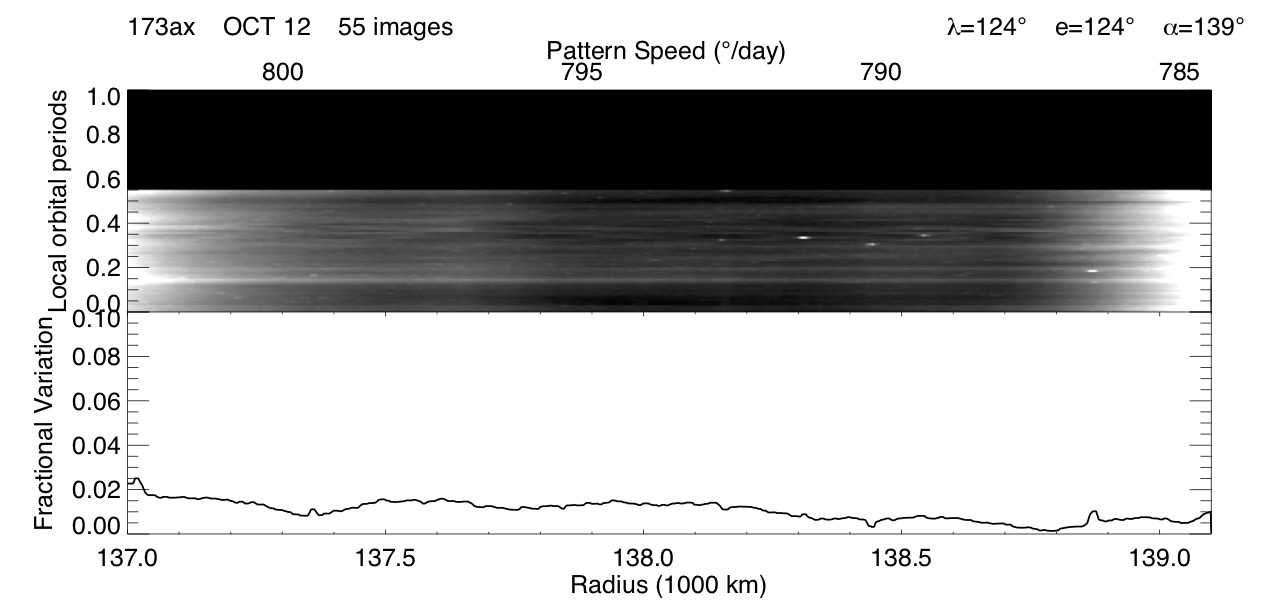}
\caption{Image movie 173ax. \label{rev173a}}
\end{figure*}

\begin{figure*}[ht]
\centering
\includegraphics[width=.76\linewidth]{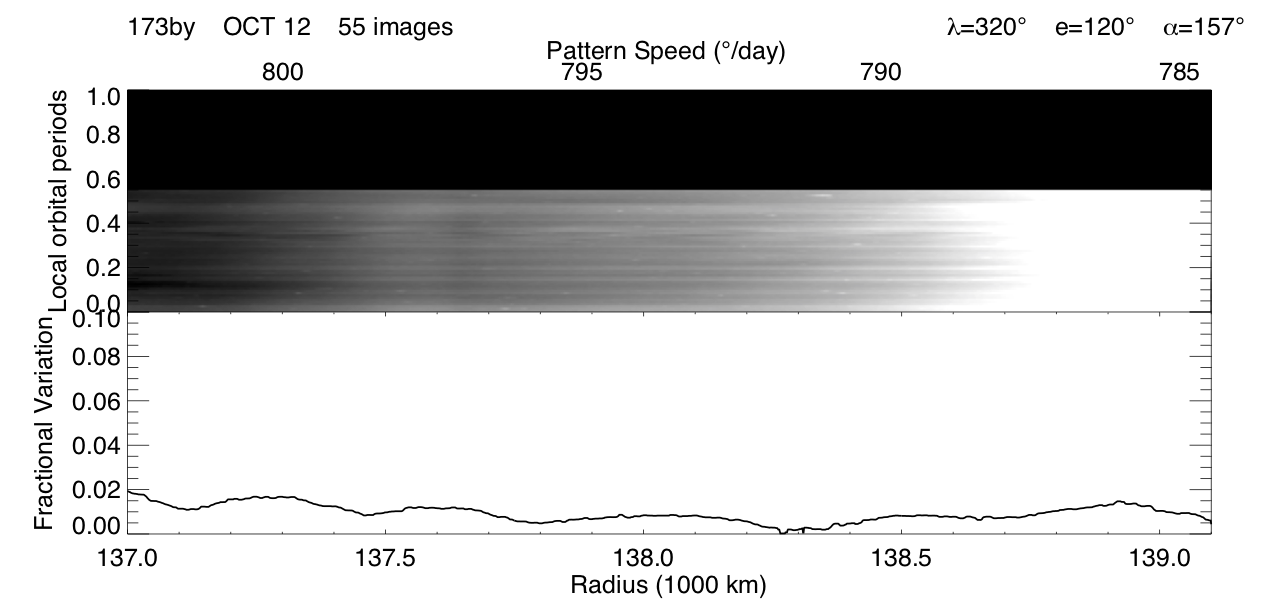}
\caption{Image movie 173by. \label{rev173b}}
\end{figure*}

\begin{figure*}[ht]
\centering
\includegraphics[width=.76\linewidth]{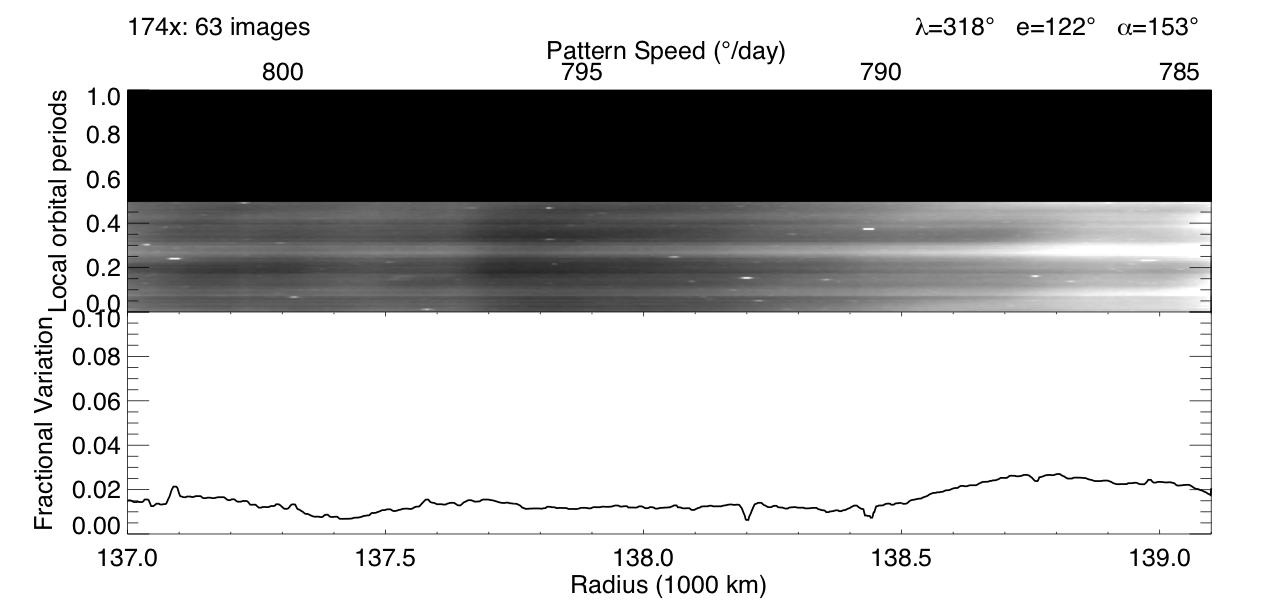}
\caption{Image movie 174x. \label{rev1741}}
\end{figure*}

\begin{figure*}[ht]
\centering
\includegraphics[width=.76\linewidth]{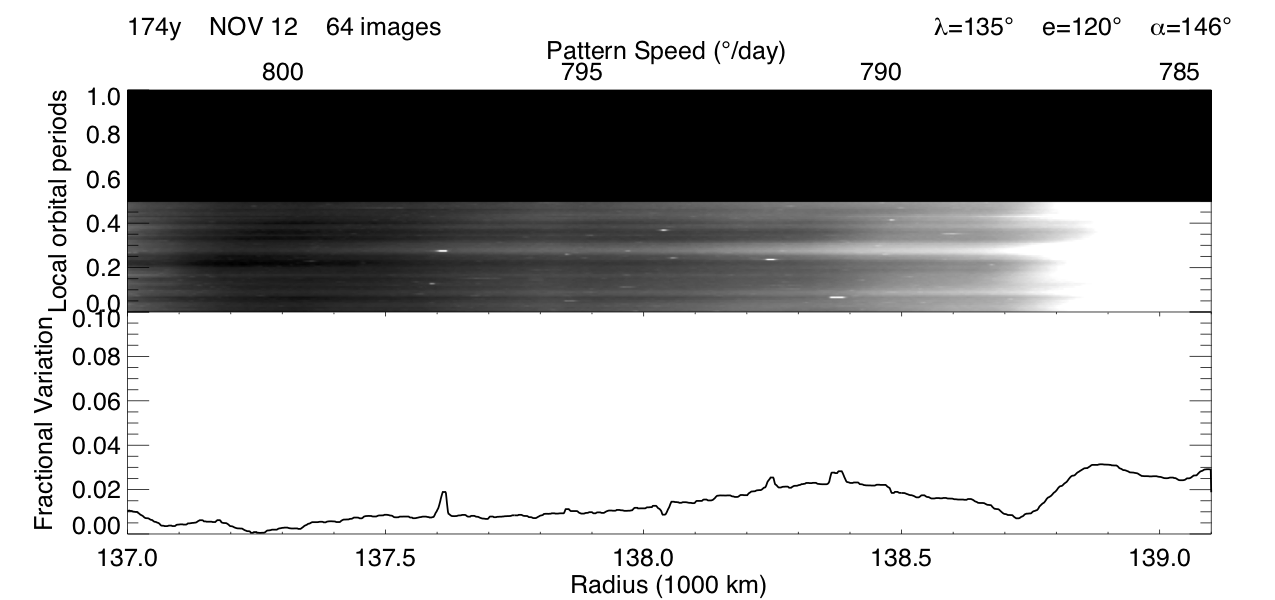}
\caption{Image movie 174y. \label{rev1742}}
\end{figure*}

\begin{figure*}[ht]
\centering
\includegraphics[width=.76\linewidth]{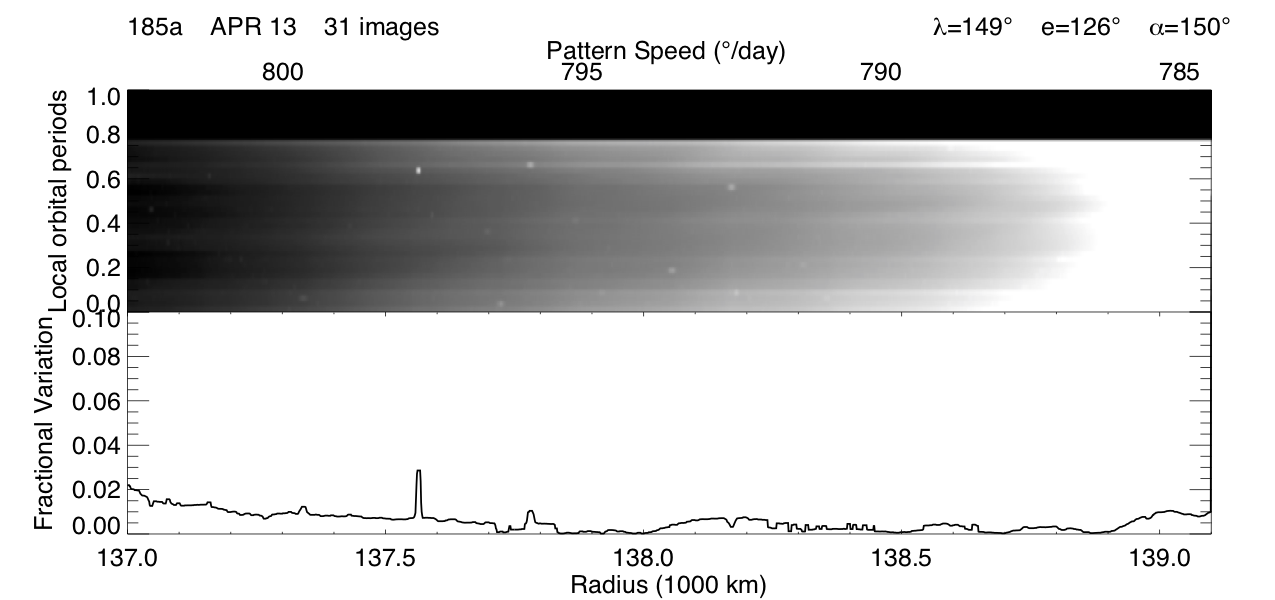}
\caption{Image movie 185a. Image movie 185b contains overexposed images due to an exposure time of $18$ seconds. \label{rev185a}}
\end{figure*}

\begin{figure*}[ht]
\centering
\includegraphics[width=.76\linewidth]{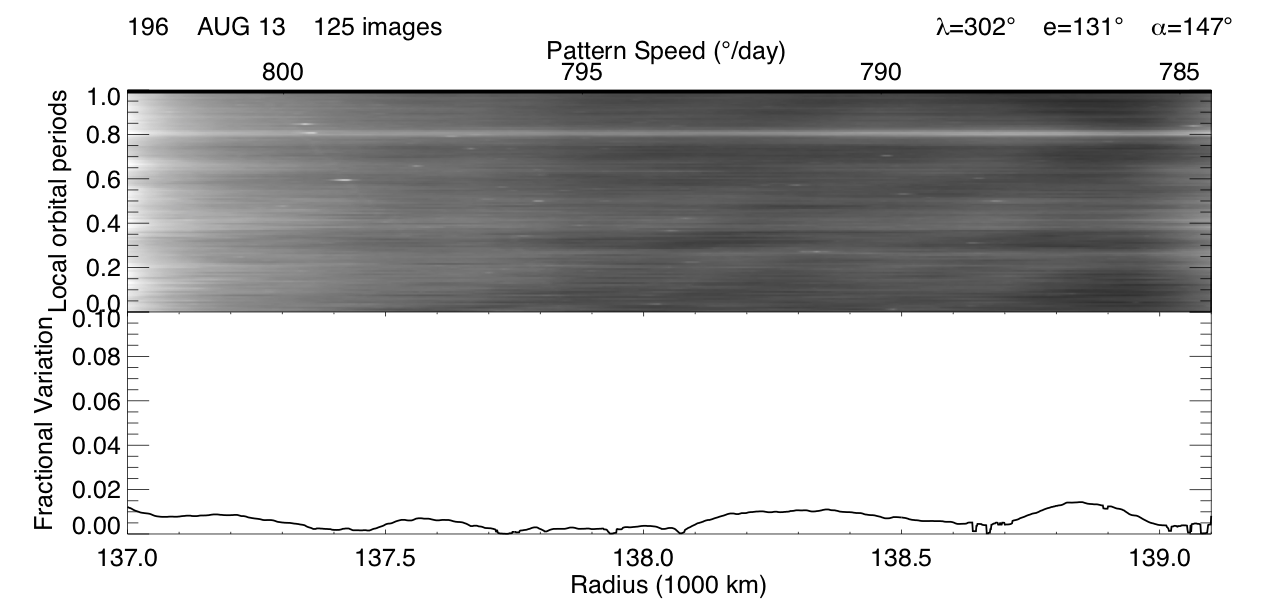}
\caption{Image movie 196. \label{rev196}}
\end{figure*}

\begin{figure*}[ht]
\centering
\includegraphics[width=.76\linewidth]{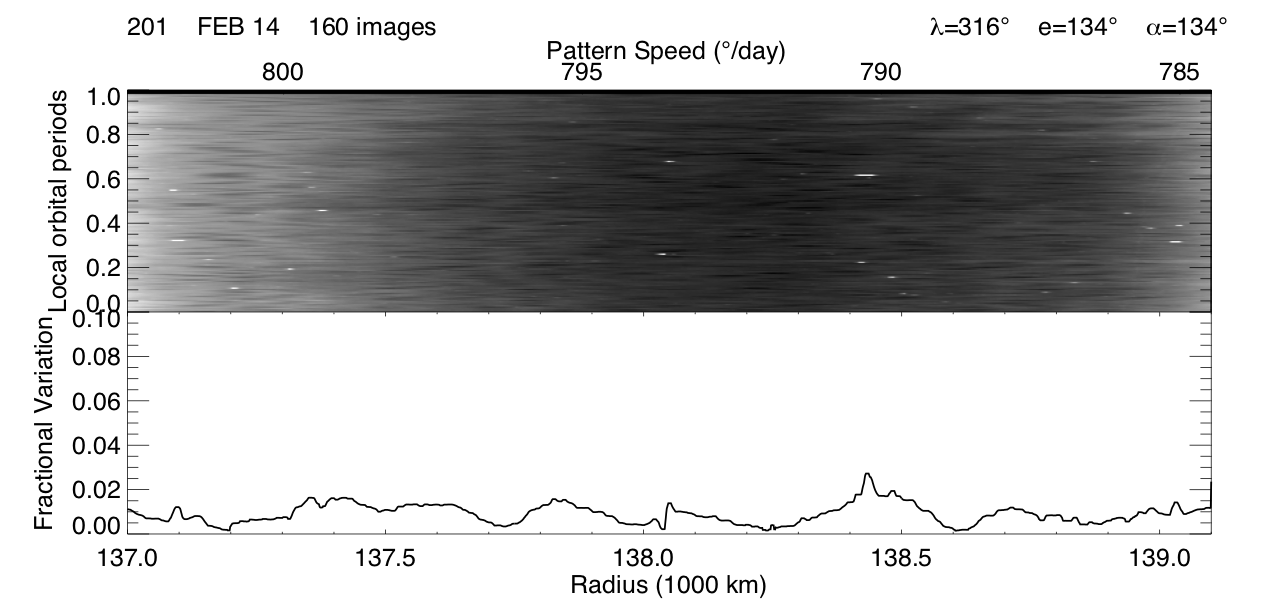}
\caption{Image movie 201. \label{rev201}}
\end{figure*}

\begin{figure*}[ht]
\centering
\includegraphics[width=.76\linewidth]{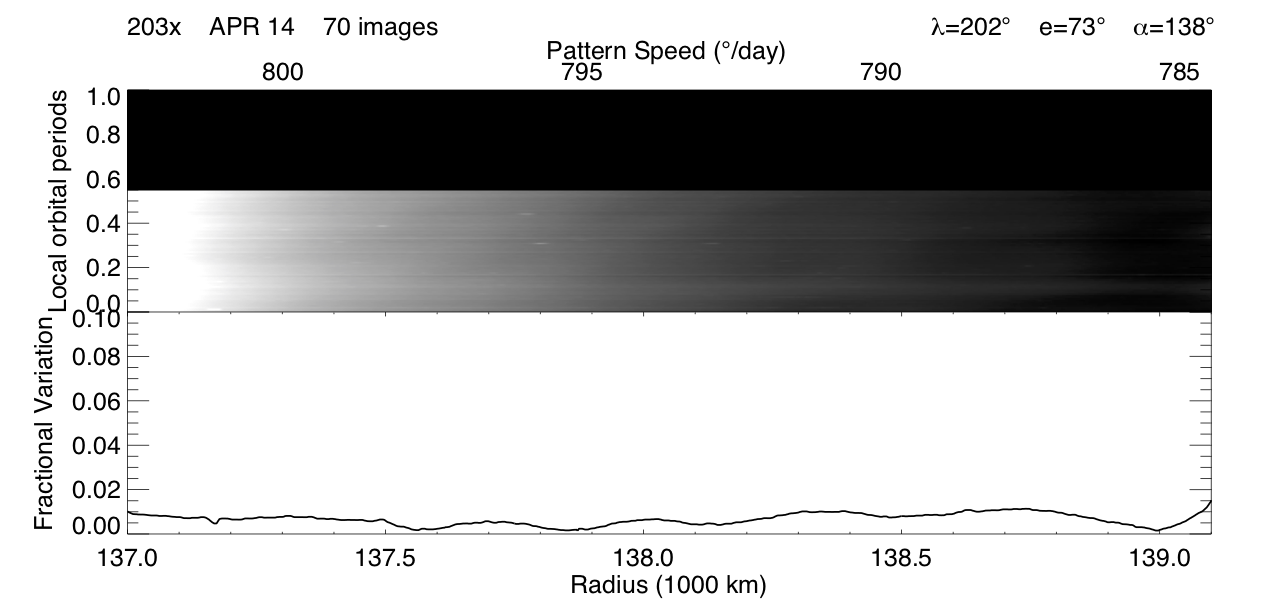}
\caption{Image movie 203x. \label{rev2031}}
\end{figure*}

\begin{figure*}[ht]
\centering
\includegraphics[width=.76\linewidth]{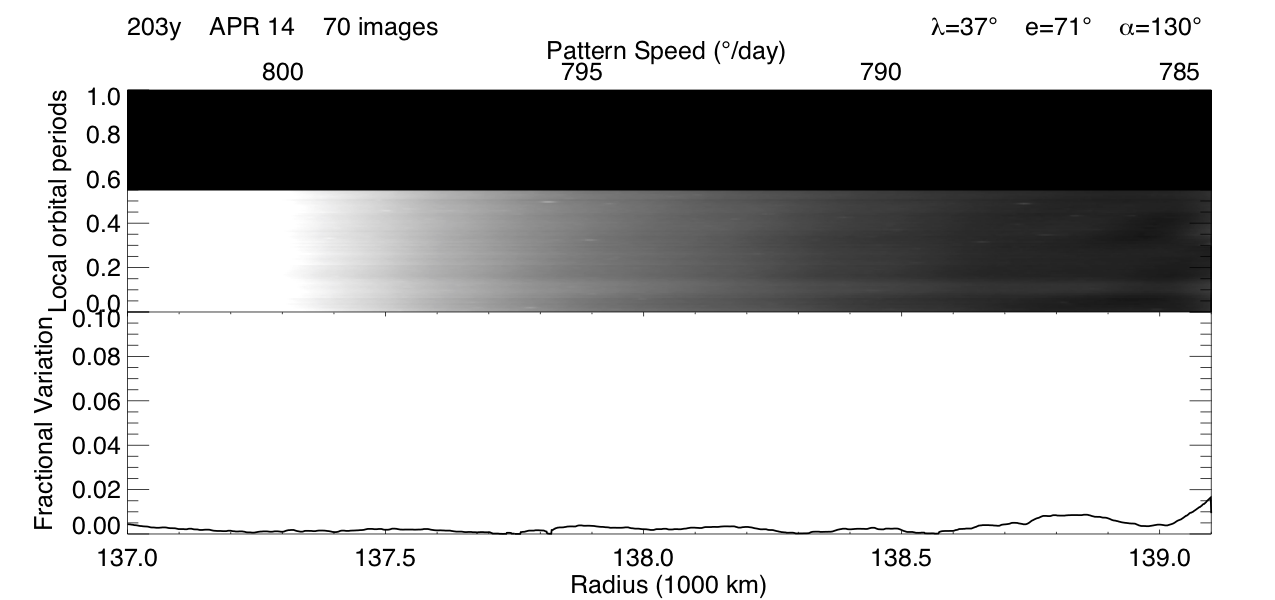}
\caption{Image movie 203y. \label{rev2032}}
\end{figure*}

Below are maps and fractional amplitude curves, equivalent to Figure \ref{example fracamp} in the main text, for all imaging sequences with durations of 7 hours or longer (including Rev 237). Fits to the fractional amplitude peaks compiled in Table \ref{patspeeds} are shown in their appropriate figures with dashed red lines. In some cases, we've combined observations with different exposure times into a single map (for example 029ax, 029bx, and 029cx into 029x). Note that in Figures \ref{rev173a} through \ref{rev2032} the SKR and MAG periods for both hemispheres fall in the A ring, well outside the radial scale shown.

\begin{figure*}[ht]
\centering
\includegraphics[width=.76\linewidth]{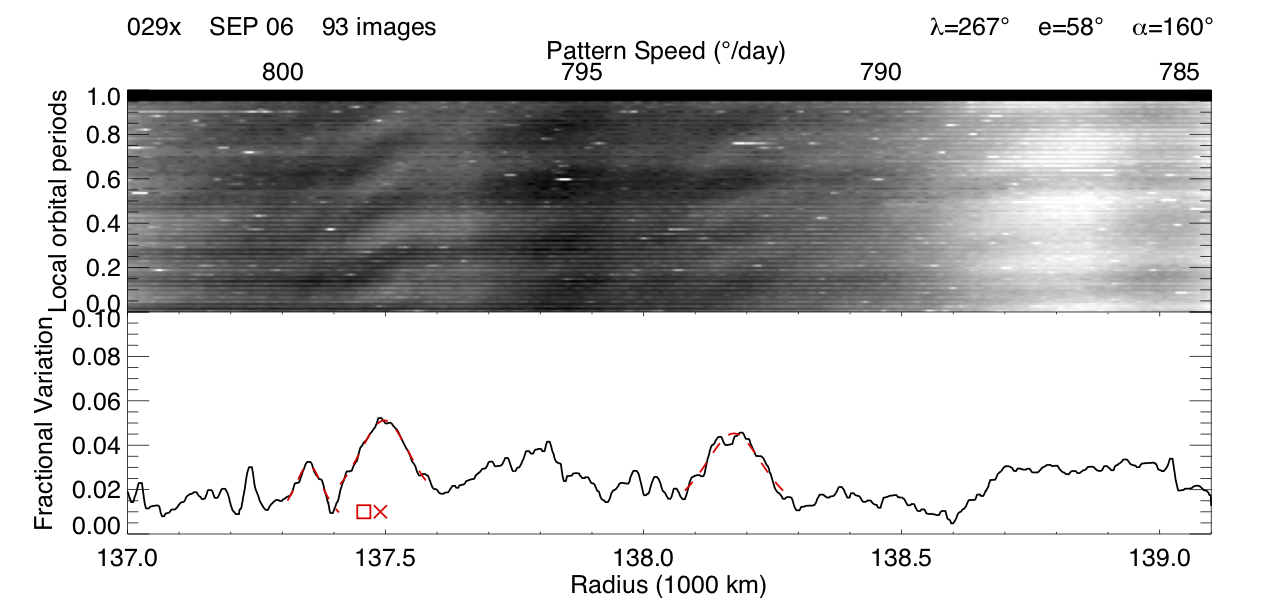}
\caption{Image movie 029x contains images from 029ax, 029bx, and 029cx. We've included the images' average inertial longitude $\lambda$, emission angle $e$, and phase angle $\alpha$ in the upper right corner. All fits to fractional amplitude peaks compiled in Table \ref{patspeeds} are shown with dashed red lines. \label{rev029x}}
\end{figure*}

\begin{figure*}[ht]
\centering
\includegraphics[width=.76\linewidth]{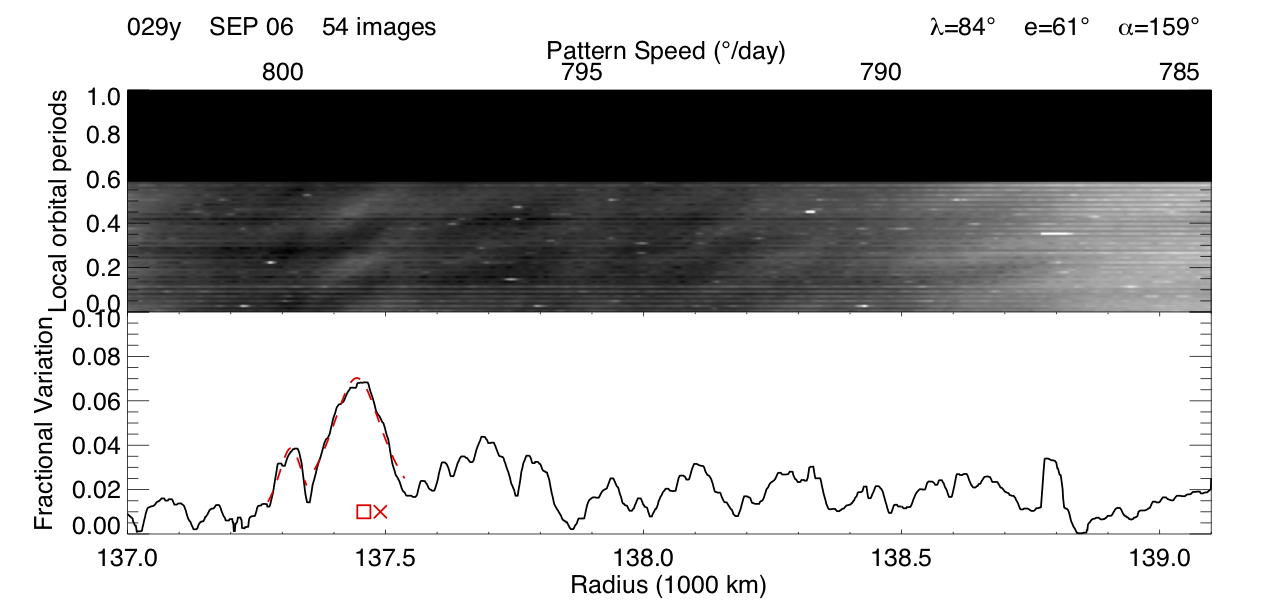}
\caption{Image movie 029y contains images from 029ay, 029by, and 029cy. \label{rev0292}}
\end{figure*}

\begin{figure*}[ht]
\centering
\includegraphics[width=.76\linewidth]{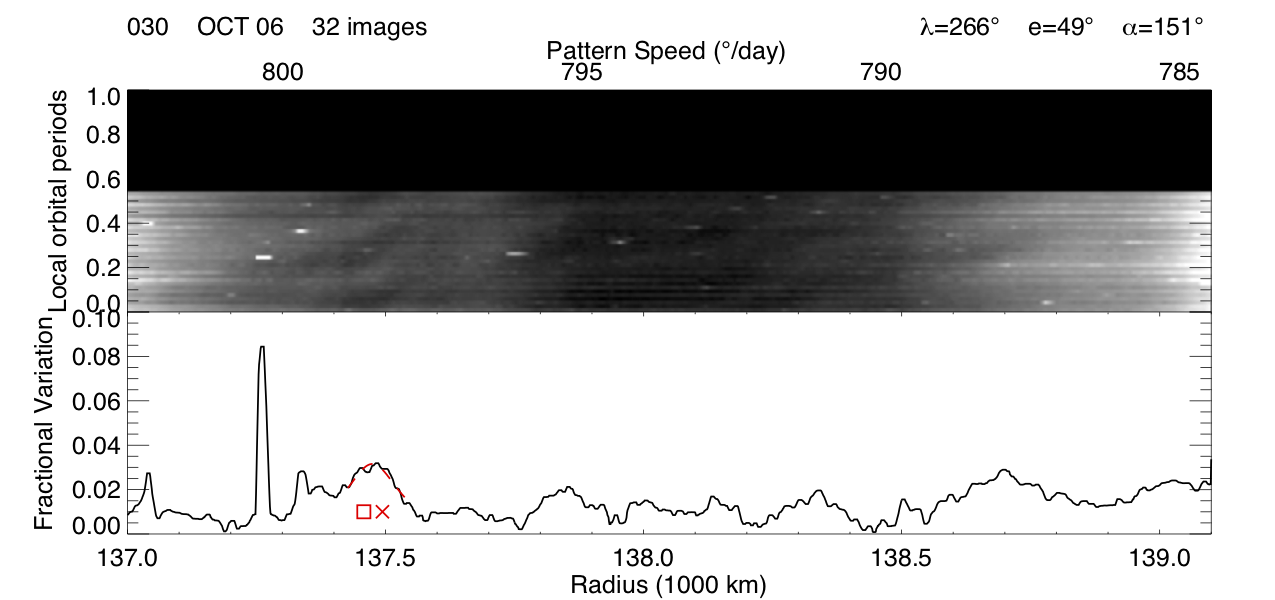}
\caption{Image movie 030 contains images from 030a, 030b, and 030c. \label{rev030}}
\end{figure*}

\begin{figure*}[ht]
\centering
\includegraphics[width=.76\linewidth]{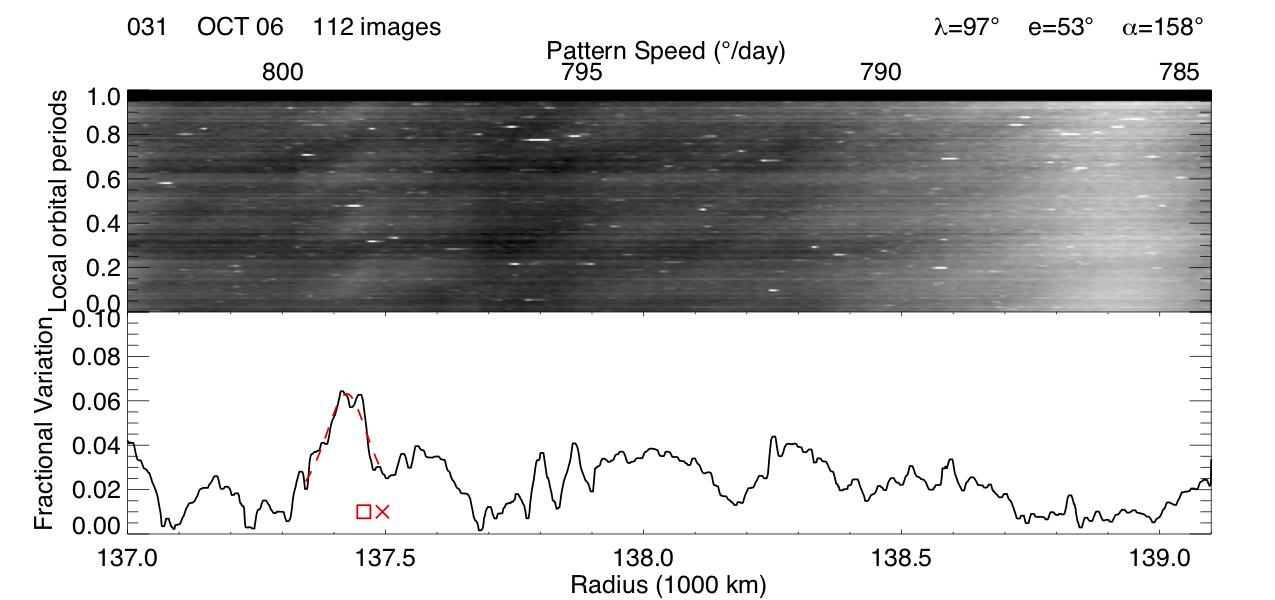}
\caption{Image movie 031 contains images from 031a, 031b, and 031cx. \label{rev031}}
\end{figure*}

\begin{figure*}[ht]
\centering
\includegraphics[width=.76\linewidth]{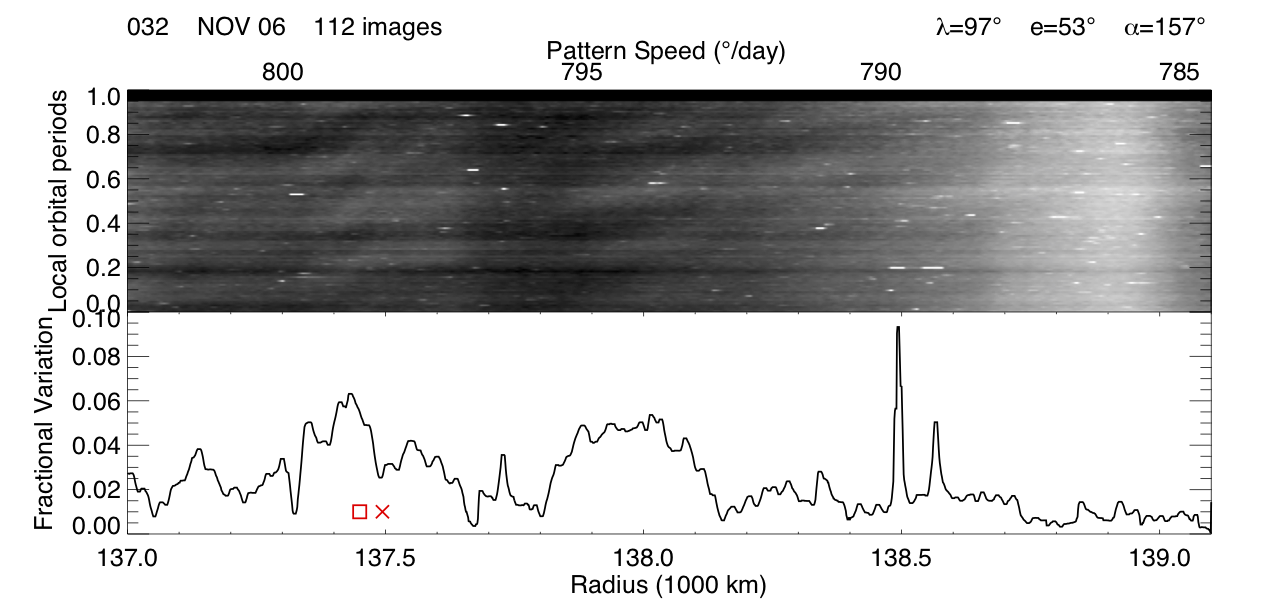}
\caption{Image movie 032 contains images from 032a, 032b, and 032c. \label{rev032}}
\end{figure*}

\begin{figure*}[ht]
\centering
\includegraphics[width=.76\linewidth]{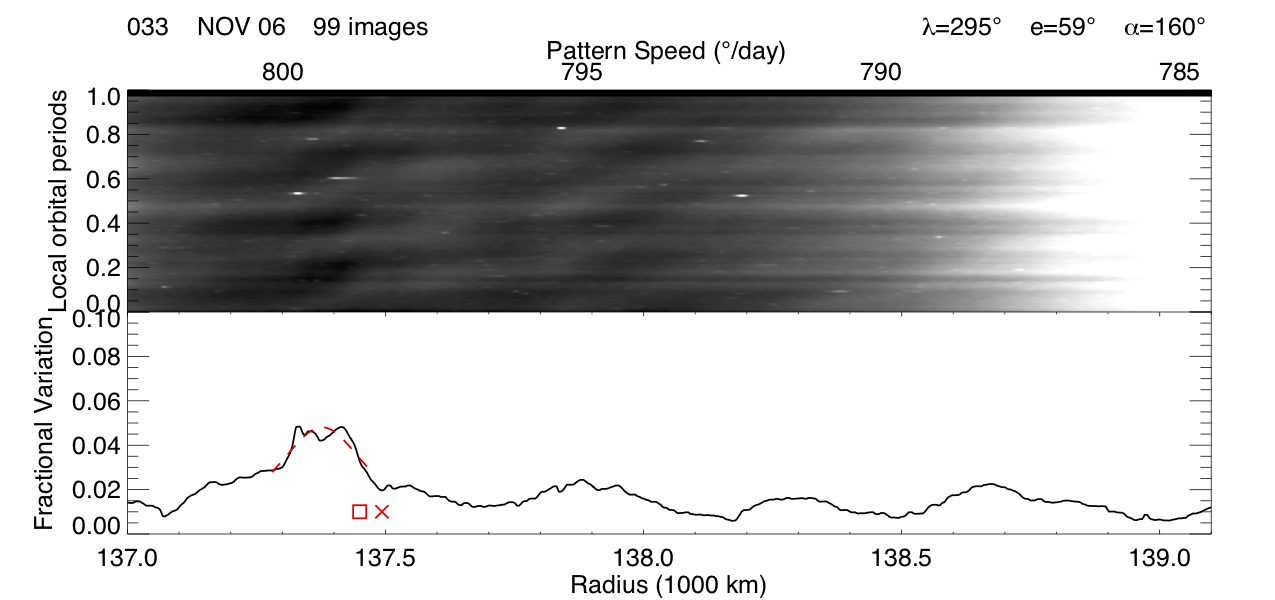}
\caption{Image movie 033. \label{rev033}}
\end{figure*}

\begin{figure*}[ht]
\centering
\includegraphics[width=.76\linewidth]{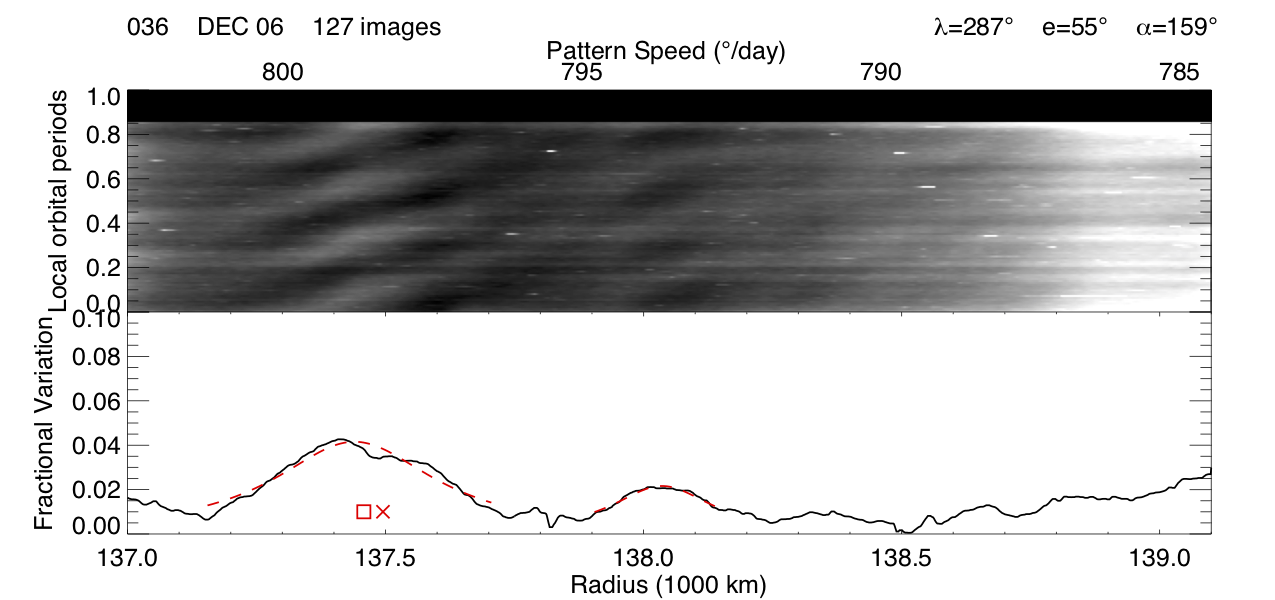}
\caption{Image movie 036. \label{rev036}}
\end{figure*}

\begin{figure*}[ht]
\centering
\includegraphics[width=.76\linewidth]{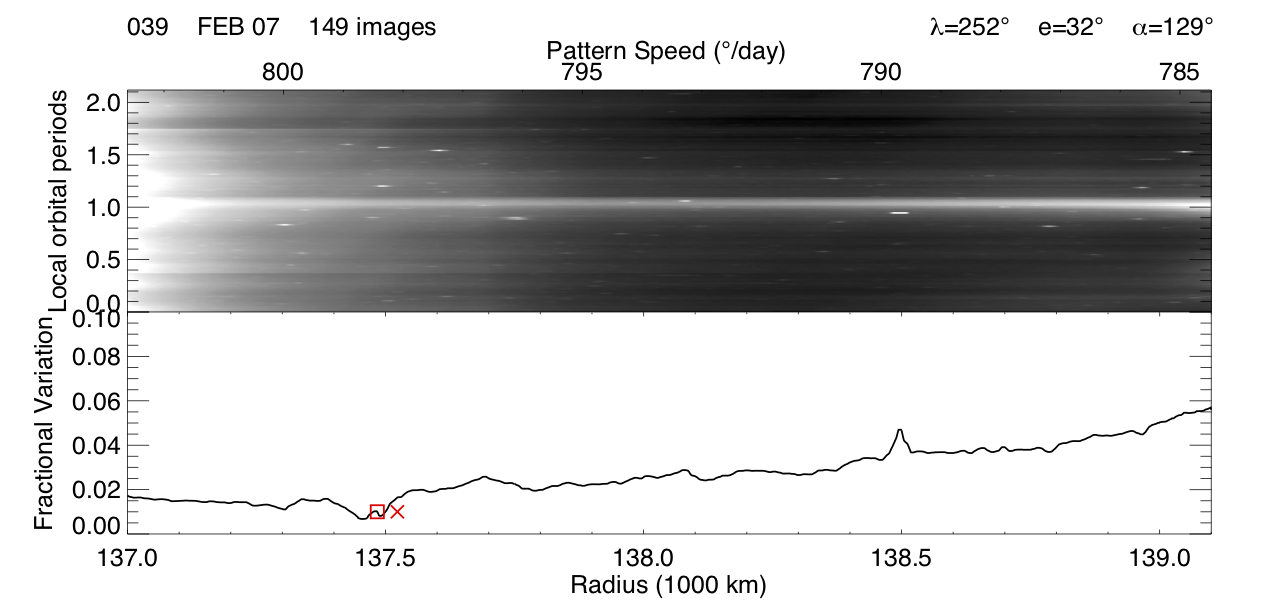}
\caption{Image movie 039 contains images from 039a and 039b. \label{rev039}}
\end{figure*}

\begin{figure*}[ht]
\centering
\includegraphics[width=.76\linewidth]{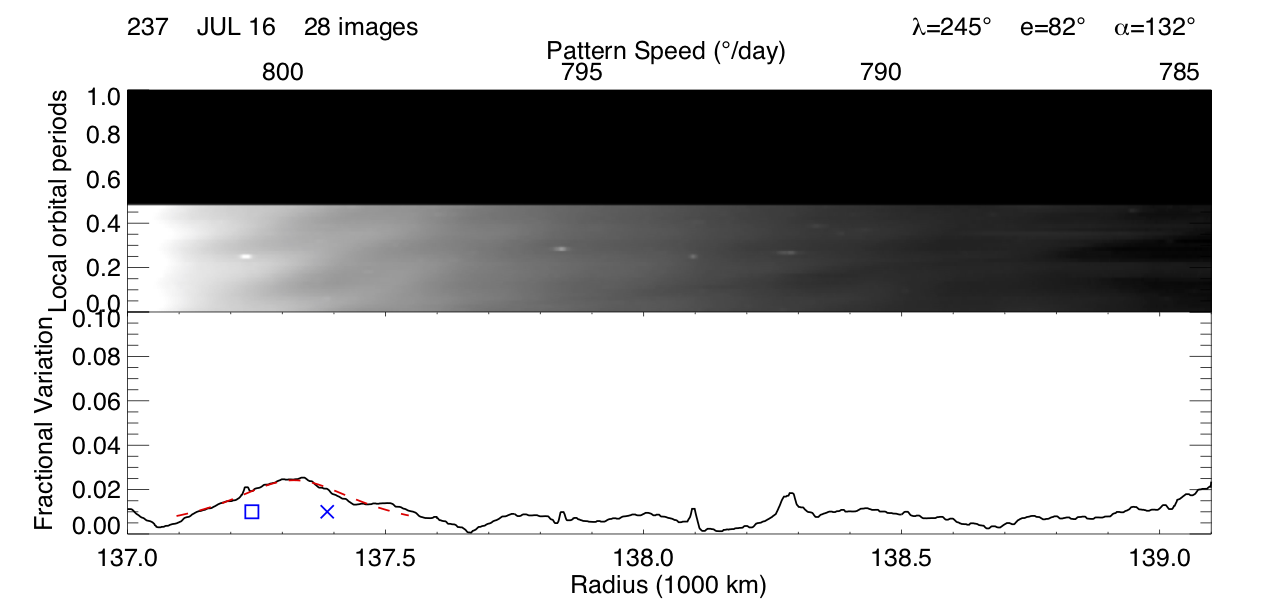}
\caption{Image movie 237. This movie is shorter than half an orbital period, but is included because it is the first reappearance of the 3:4 OLR structures in the Roche Division. \label{rev237}}
\end{figure*}

\begin{figure*}[ht]
\centering
\includegraphics[width=.76\linewidth]{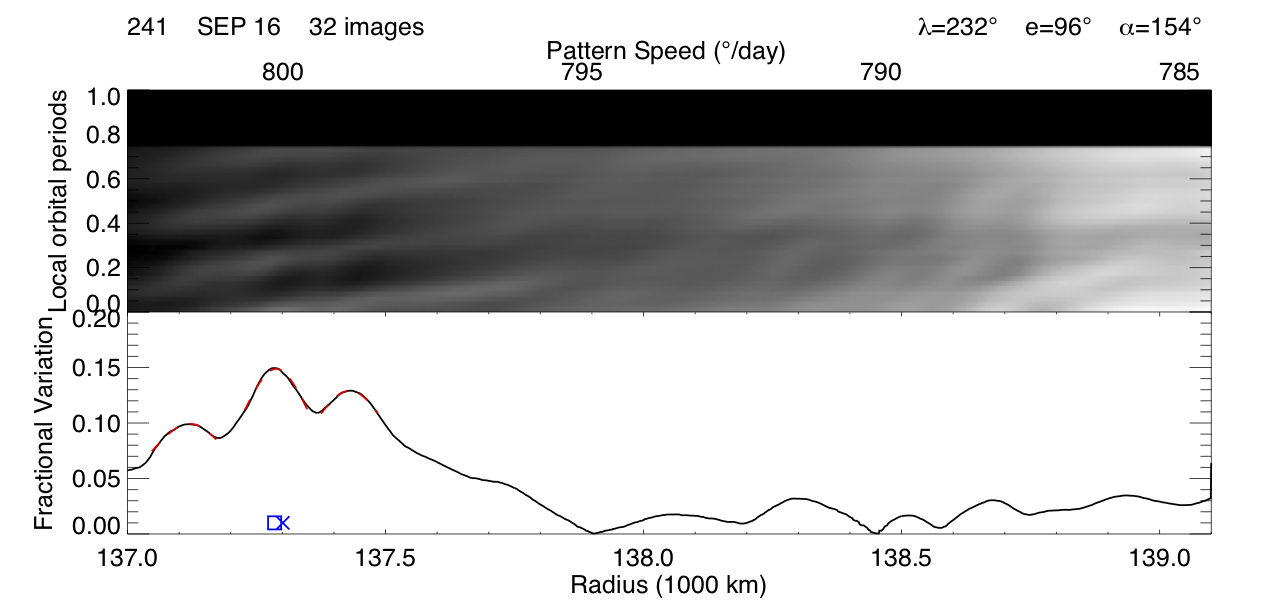}
\caption{Image movie 241. The scale on the fractional variation plot is increased in this plot alone. The brightness variations are substantially enhanced by the favorable lighting geometry of this observation. \label{rev241}}
\end{figure*}

\begin{figure*}[ht]
\centering
\includegraphics[width=.76\linewidth]{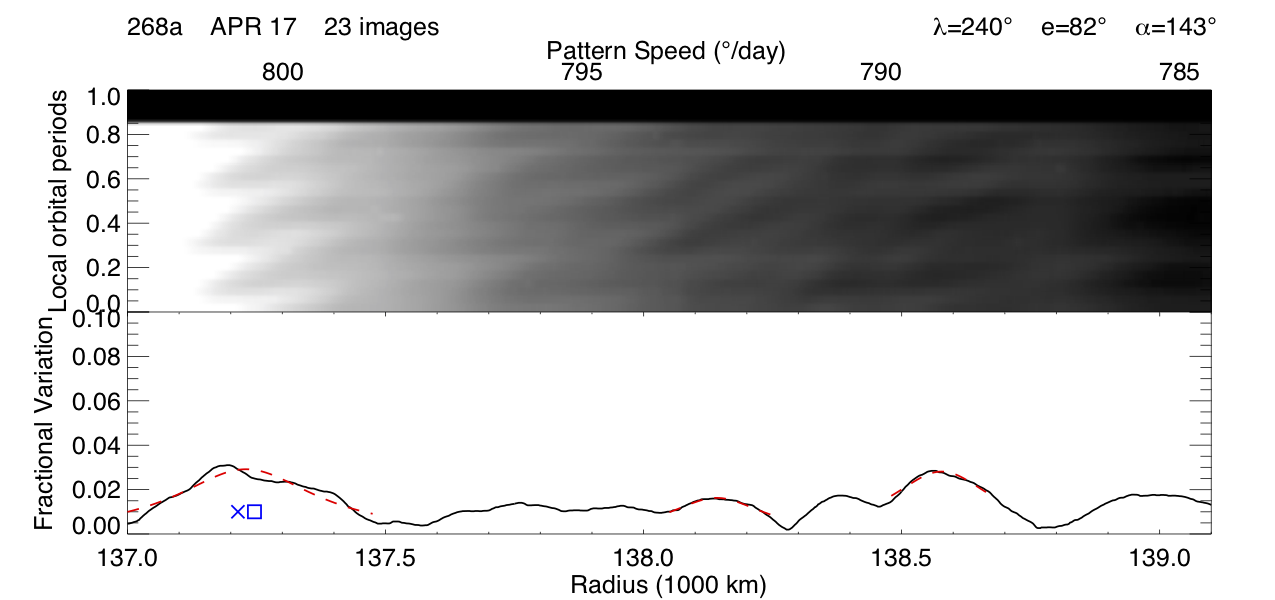}
\caption{Image movie 268a. Image movie 268b contains overexposed images due to an exposure time of $18$ seconds. \label{rev268a}}
\end{figure*}

\begin{figure*}[ht]
\centering
\includegraphics[width=.76\linewidth]{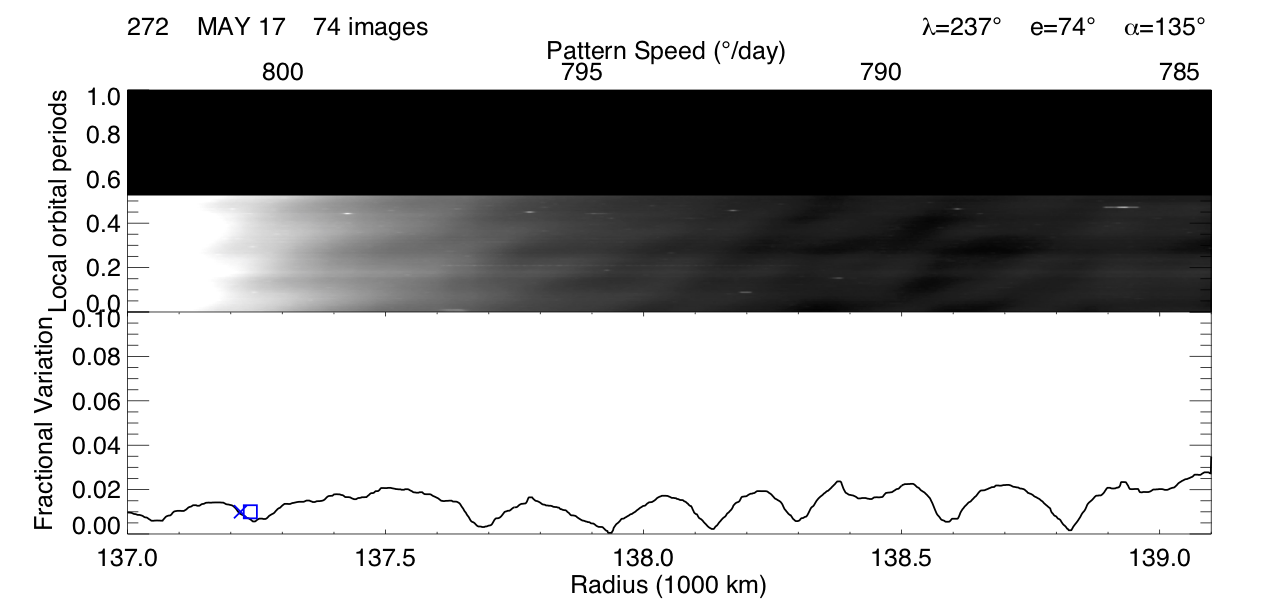}
\caption{Image movie 272. \label{rev272}}
\end{figure*}

\begin{figure*}[ht]
\centering
\includegraphics[width=.76\linewidth]{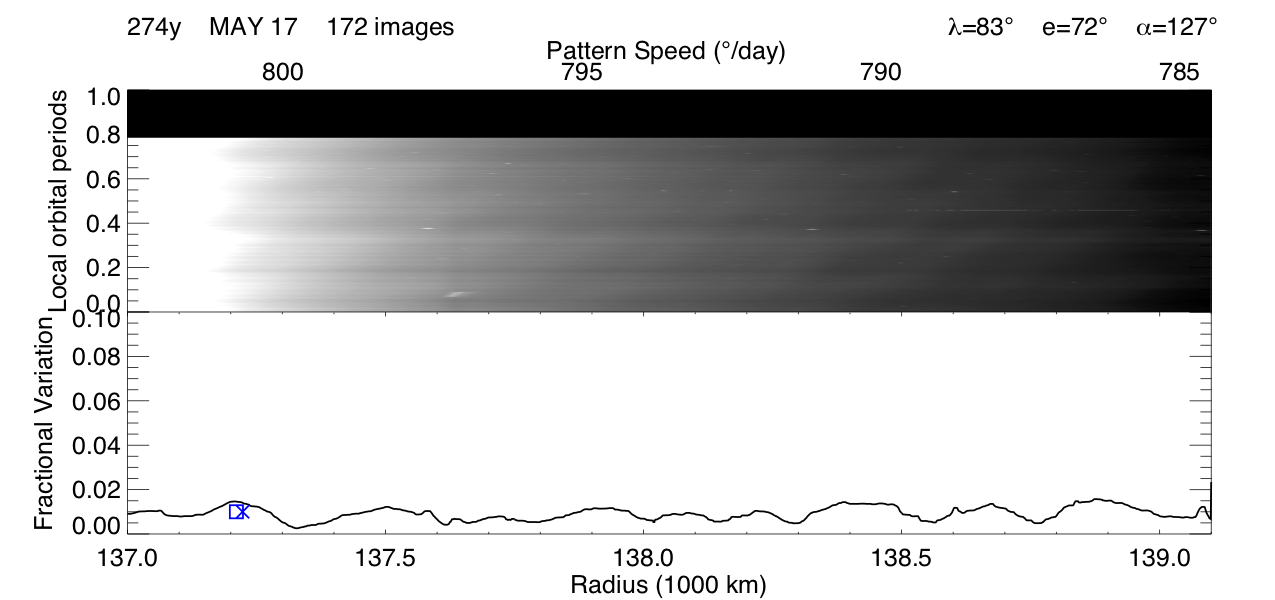}
\caption{Image movie 274y. \label{rev2742}}
\end{figure*}

\begin{figure*}[ht]
\centering
\includegraphics[width=.76\linewidth]{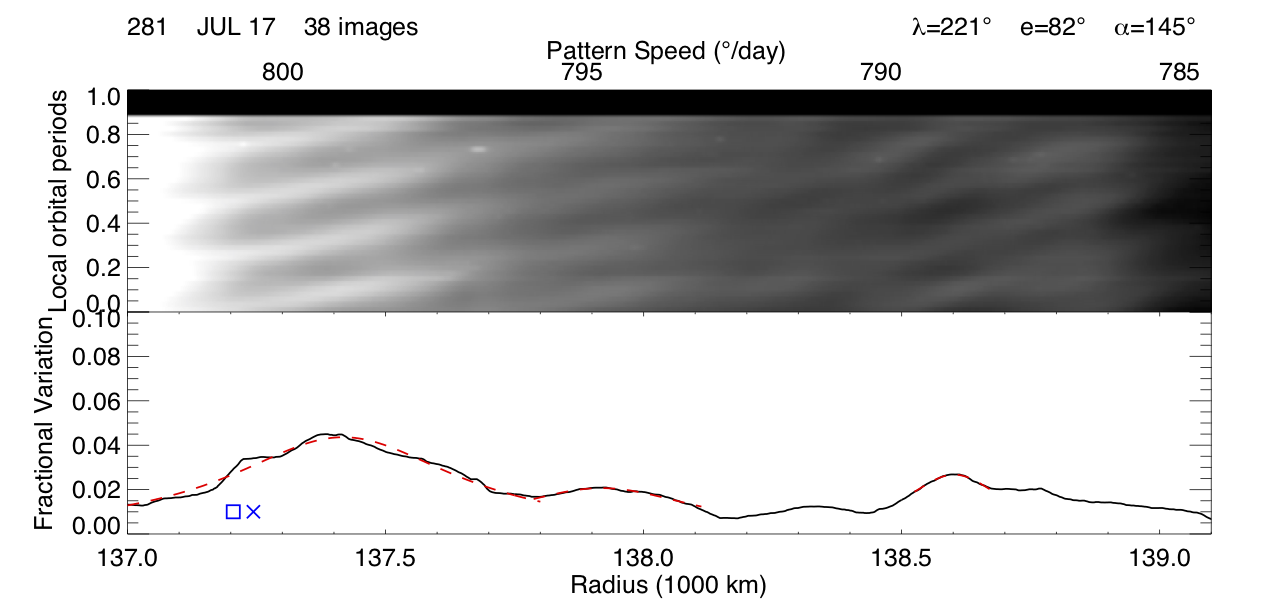}
\caption{Image movie 281. \label{rev281}}
\end{figure*}

\begin{figure*}[ht]
\centering
\includegraphics[width=.76\linewidth]{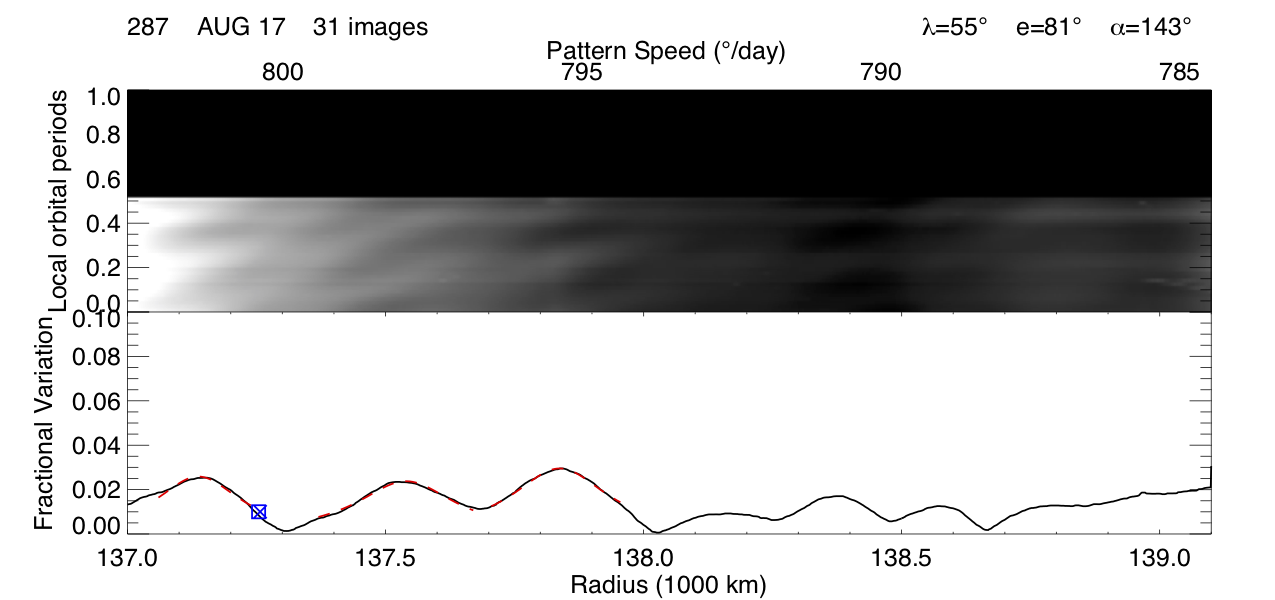}
\caption{Image movie 287. \label{rev287}}
\end{figure*}

\begin{figure*}[ht]
\centering
\includegraphics[width=.76\linewidth]{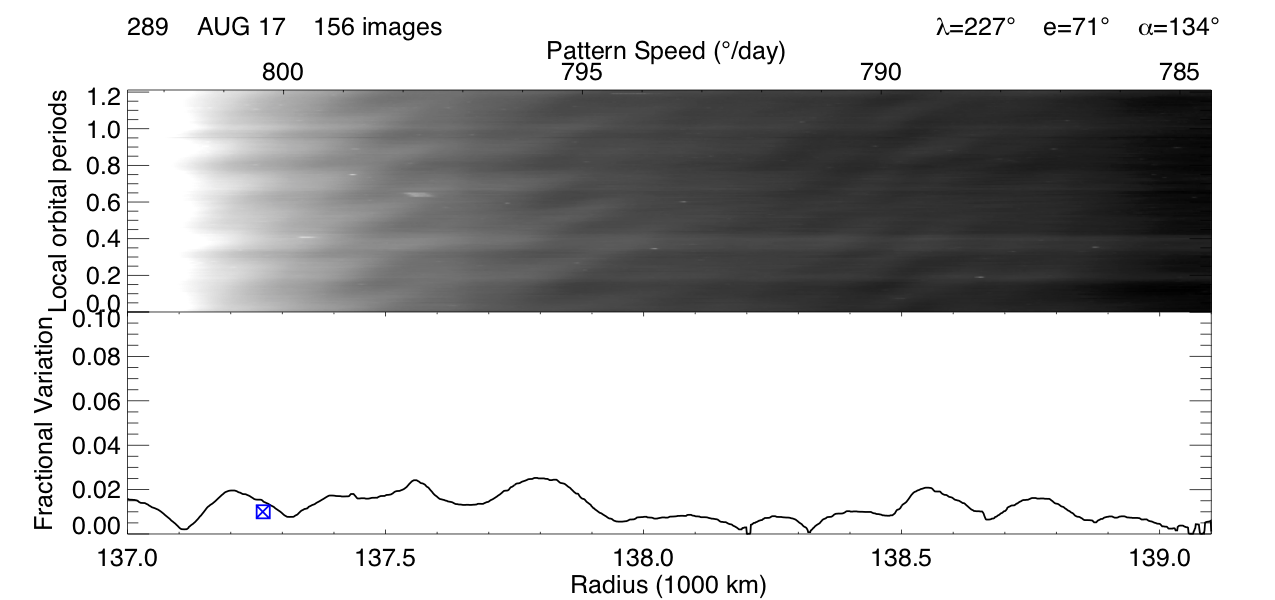}
\caption{Image movie 289. \label{rev289}}
\end{figure*}

\begin{figure*}[ht]
\centering
\includegraphics[width=.76\linewidth]{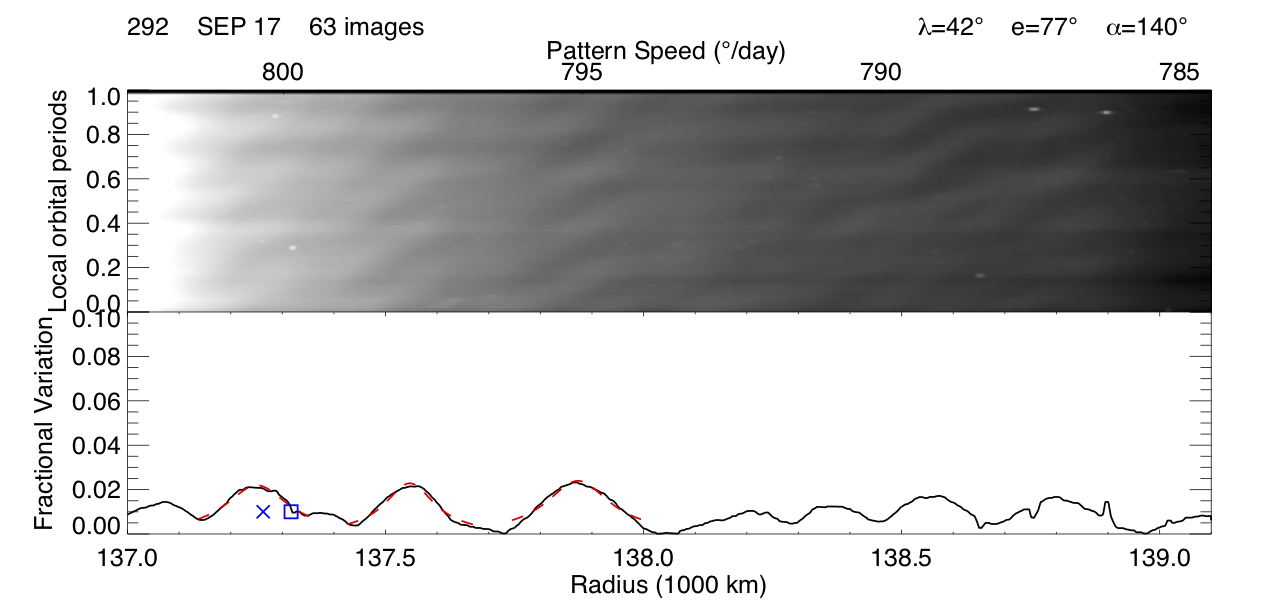}
\caption{Image movie 292. \label{rev292}}
\end{figure*}


\end{document}